\documentclass[draftclsnofoot, onecolumn, letterpaper,
romanappendices]{IEEEtran}
\usepackage{Albert_Style}
\usepackage{bookmark}% faster updated bookmarks
\begin{document}
\title{Strong Successive Refinability and Rate-Distortion-Complexity Tradeoff}

\author{
\IEEEauthorblockN{Albert No\IEEEauthorrefmark{1}, Amir Ingber\IEEEauthorrefmark{2}, Tsachy Weissman\IEEEauthorrefmark{1}\\}
\IEEEauthorblockA{\IEEEauthorrefmark{1}Department of Electrical Engineering, Stanford University\\
\texttt{\{albertno, tsachy\}@stanford.edu}}\\
\IEEEauthorblockA{\IEEEauthorrefmark{2}Yahoo! Labs\\
\texttt{Ingber@yahoo-inc.com}}
\thanks{The material in this paper has been presented in part at the 2013 51st Annual Allerton Conference on Communication, Control, and Computing (Allerton), and at the 2014 International Symposium on Information Theory.  This work was supported by the NSF Center for Science of Information under Grant Agreement CCF-0939370.}
}

\maketitle

%%%%%%%%%%%%%%%%%%%%%%%%%%%%%%%%%%%%%%%%%%%%%%%%
%%%%%%%%%%%%%%%%           Abstract                           %%%%%%%%%%%%%%%%%%
%%%%%%%%%%%%%%%%%%%%%%%%%%%%%%%%%%%%%%%%%%%%%%%%
\begin{abstract}
We investigate the second order asymptotics (source dispersion) of the successive refinement problem. Similarly to the classical definition of a successively refinable source, we say that a source is strongly successively refinable if successive refinement coding can achieve the second order optimum rate (including the dispersion terms) at both decoders. We establish a sufficient condition for strong successive refinability. We show that any discrete source under Hamming distortion and the Gaussian source under quadratic distortion are strongly successively refinable. 

We also demonstrate how successive refinement ideas can be used in point-to-point lossy compression problems in order to reduce complexity. We give two examples, the binary-Hamming and Gaussian-quadratic cases, in which a layered code construction results in a low complexity scheme that attains optimal performance. For example, when the number of layers grows with the block length $n$, we show how to design an $O(n^{\log(n)})$ algorithm that asymptotically achieves the rate-distortion bound.

% Keywords
\end{abstract}
\IEEEpeerreviewmaketitle

%%%%%%%%%%%%%%%%%%%%%%%%%%%%%%%%%%%%%%%%%%%%%%%%
%%%%%%%%%%%%%%%%           Introduction                            %%%%%%%%%%%%%%%%
%%%%%%%%%%%%%%%%%%%%%%%%%%%%%%%%%%%%%%%%%%%%%%%%
\section{Introduction}\label{sec:Introduction}

 In the successive refinement problem, an encoder wishes to send a source to two decoders with different target distortions. Instead of designing separate coding schemes, the successive refinement encoder uses a code for the first decoder which has a weaker link and sends extra information to the second decoder on top of the message of the first decoder. In general, the performance of a successive refinement coding scheme is worse than separate coding for each decoder. However, for some cases, we can simultaneously achieve the optimum rates for both decoders as if the optimum codes were used separately. In this case, we say the source is successively refinable. Necessary and sufficient conditions for successive refinement were independently proposed by Kosh\'elev~\cite{koshelev1980hierarchical, koshelev1981estimation} and Equitz and Cover~\cite{equitz1991successive}. Rimoldi \cite{rimoldi1994successive} found the full rate-distortion region of the successive refinement problem including non-successively refinable sources. Kanlis and Narayan \cite{kanlis1996error} extended the result to the error exponent that quantifies ``how fast the excess distortion probability decays". Tuncel \cite{tuncel2003error} characterized the entire region of rate-distortion-exponents with separate handling of the two error events. Both lines of work considered error exponents in the spirit of Marton \cite{marton1974error}, which characterized the error exponent for the point-to-point case.

For the point-to-point source coding problem, Ingber and Kochman \cite{ingber2011dispersion} and Kostina and Verd\`u \cite{kostina2012fixed} independently proposed an asymptotic analysis that complements the error exponent analysis. In this setting, the figure of merit is the minimum achievable rate when the excess distortion probability $\epsilon$ and the block length $n$ are fixed. This can be quantified by the source dispersion. For an i.i.d.\ source with law $P$, the minimum rate can be approximated by $R(P,D)+ \sqrt{{V(P,D)}/{n}}Q^{-1}(\epsilon)$, where $R(P,D)$ and $V(P,D)$ are, respectively, the rate-distortion function and dispersion of a source $P$ at distortion level $D$. We can consider this rate as a ``second order'' optimum rate (where the classical rate-distortion function is the first order result).

With this stronger notion of optimality, it is natural to ask whether successive refinement schemes can achieve the second order optimum rates at both decoders simultaneously. An obvious necessary condition for the existence of such schemes is that the source be successively refinable, so we refer to such a source as ``strongly successively refinable" (formal definitions follow in the sequel). In this paper, we present a second order achievability result for the successive refinement problem. As a corollary, we derive a sufficient condition for strong successive refinability and show that a source $P$ is strongly successively refinable if all sources $\tilde{P}$ in the neighborhood of $P$ are successively refinable.

In the second part of the paper, we show that successive refinement codes can be useful in the point-to-point source coding problem when we want to achieve lower encoding complexity. The idea is that finding the best representing codeword in a successive manner is often easier than finding a codeword from the set of all codewords, which normally has exponential complexity. Moreover, storing exponentially many codewords is often prohibitive, while successive refinement encoding can reduce the size of codebooks. Our findings here contribute to the recent line of work on reducing the complexity of rate-distortion codes, cf.\ \cite{gupta2008rate, korada2010polar, venkataramanan2014lossy} and references therein. 

We aim to study the general approach of using successive encoding to reduce complexity. We denote this approach by ``layered coding", a family that includes all coding schemes that can be implemented in a successive manner. Basically, the layered coding scheme is searching for an appropriate codeword over a tree structure where the number of decoders corresponds to the level of the tree. The larger the tree, the faster the codeword can be found, and therefore the lower decoding complexity. In order to reduce the encoding complexity significantly, we generalize the result to the case where the number of decoders is increasing with block length $n$. This is different from the classical successive refinability where only a fixed number of decoders are considered. On the other hand, the larger tree structure restricts the class of coding schemes, and therefore too many decoders may cause a rate loss. Our result for this setting characterizes an achievable trade-off between encoding complexity (how fast can we find the codeword) and performance (how much do we end up compressing). Note that SPARC \cite{venkataramanan2014lossy} and CROM \cite{no2014rateless} are manifestations of the layered coding approach that attain good performance. 

The rest of the paper is organized as follows. In Section \ref{sec:Preliminaries}, we revisit the known results about successive refinement and source dispersion. Section \ref{sec:Problem Setting} provides the problem setting. We present our main results in Section \ref{sec:Main Results}, where proof details are given in Section \ref{sec:Proof}. Section \ref{sec:Layered Codes} is dedicated to a layered coding scheme, and we conclude in Section \ref{sec:Conclusions}.

\emph{Notation}: $X^n$ and $\bX$ denotes an $n$-dimensional random vector $(X_1,X_2,\ldots,X_n)$ while $x^n$ and $\bx$ denotes a specific realization of it. When we have two random vectors, we use the notation such as $\Xh_1^n = (\hat{X}_{1,1},\hat{X}_{1,2},\ldots,\hat{X}_{1,n})$ and $\Xh_2^n = (\hat{X}_{2,1},\hat{X}_{2,2},\ldots,\hat{X}_{2,n})$.

%%%%%%%%%%%%%%%%%%%%%%%%%%%%%%%%%%%%%%%%%%%%%%%%
%%%%%%%%%%%%%%%%           Preliminaries                            %%%%%%%%%%%%%%%%
%%%%%%%%%%%%%%%%%%%%%%%%%%%%%%%%%%%%%%%%%%%%%%%%
\section{Preliminaries}\label{sec:Preliminaries}

%%%%%%%%%%%%%%%%           Source Dispersion
%%%%%%%%%%%%%%%%%%%%%%%%%%%%%%%%%%%%%%%%%%%%%%%%
\subsection{Source Dispersion}\label{subsec:Source Dispersion}
Consider an i.i.d.\ source $X^n$ with law $P$ where the source alphabet is $\cX$ and the reconstruction alphabet is $\cXh $. Let $d:\cX\times\cXh \rightarrow[0,\infty)$ be a distortion measure where $d(x^n,\hat{x}^n) = (1/n) \sum_{i=1}^n d(x_i,\hat{x}_i)$. It is well known that the rate-distortion function $R(P,D)$ is the optimal asymptotic compression rate for which distortion $D$ can be achieved. However, this first order optimum rate can be achieved only when the block length $n$ goes to infinity. Beyond the first order rate, we can consider two\footnote{These asymptotic approaches analyze the \emph{excess distortion probability}. Other approaches exist which analyze the \emph{average achievable distortion} \cite{yang1999redundancy, zhang1997redundancy}.} asymptotic behaviors which are excess distortion exponent \cite{marton1974error} and the source dispersion \cite{DBLP:journals/corr/abs-1109-6310, kostina2012fixed}. The former considers how fast the excess distortion probability $\Pr{d(X^n,\hat{X}^n)>D}$ is decaying, while the latter considers how fast the minimum number of codewords converges to $R(P,D)$ when excess distortion probability $\epsilon$ and block length $n$ are given. It was shown that the difference between the minimum rate for fixed $n$ and $R(P,D)$ is inversely proportional to square root of $n$. More formally, let $R_{P,D,\epsilon}(n)$ be the minimum compression rate for which the excess distortion probability is smaller than $\epsilon$. The result is given by:
\begin{theorem}[\cite{DBLP:journals/corr/abs-1109-6310}]\label{thm:IngberDispersion}
Suppose $R(P,D)$ is twice differentiable\footnote{We say $R(P, D)$ is differentiable at $P$ if there is an extension $\tilde{R}(\cdot,D):\fR^m \rightarrow \fR$ which is differentiable. Under this definition, $R'(x,D)$ and $V(P,D)$ are well and uniquely defined. Details are given in Appendix \ref{app:Derivative of Rate-Distortion Function}.} with respect to $D$ and the elements of $P$ in some neighborhood of $(P,D)$. Then
\begin{align}
R_{P,D,\epsilon}(n) = R(P,D) + \sqrt{\frac{V(P,D)}{n}}Q^{-1}(\epsilon) + O\left(\frac{\log n}{n}\right)
\end{align}
where $V(P,D)$ is the \emph{source dispersion}, given by
\begin{align}
V(P,D) \triangleq& \Var{R'(X,D)} \\
=& \sum_{x\in\cX} P(x)(R'(x,D))^2 - \left[\sum_{x\in\cX} P(x)R'(x,D)\right]^2
\end{align}
and $R'(x,D)$ denotes the derivative of $R(P,D)$ with respect to the probability $P(x)$:
\begin{align}
R'(x,D)\triangleq \left[\frac{\partial R(Q,D)}{\partial Q(x)}\right]_{Q=P}.
\end{align}
\end{theorem}

We have a similar result for the Gaussian source under quadratic distortion:
\begin{theorem}[\cite{ingber2011dispersion}]\label{thm:IngberDispersionGaussian}
Consider an i.i.d.\ Gaussian source $X^n$ distributed according to $\cN(0,\sigma^2)$, and quadratic distortion, i.e., $d(x^n, \hat{x}^n) = (1/n) \sum_{i=1}^n (x_i-\hat{x}_i)^2$. Then
\begin{align}
R_{P,D,\epsilon}(n) = \frac{1}{2}\log \frac{\sigma^2}{D} + \sqrt{\frac{1}{2n}}Q^{-1}(\epsilon) + O\left(\frac{\log n}{n}\right).
\end{align}
Note that the dispersion of the Gaussian source is $V(P,D)=1/2\mbox{ nats}^2/\mbox{source symbol}$ for all $D\leq \sigma^2$.
\end{theorem}

%%%%%%%%%%%%%%%%           Successive Refinement
%%%%%%%%%%%%%%%%%%%%%%%%%%%%%%%%%%%%%%%%%%%%%%%%
\subsection{Successive Refinement}\label{subsec:Successive Refinement}
The successive refinement problem with two decoders can be formulated as follows. Again, let $X^n$ be i.i.d.\ with law $P$. The encoder sends a pair of messages $(m_1,m_2)$ where $1\leq m_i\leq M_i$ for $i \in \{1,2\}$. The first decoder takes $m_1$ and reconstructs $\Xh_1^n(m_1)\in\cXh_1^n$  where the second decoder takes $(m_1,m_2)$ and reconstructs $\Xh_2^n(m_1,m_2)\in\cXh_2^n$. Note that $\cXh_1$ and $\cXh_2$ denote the respective reconstruction alphabets of the decoders. The $i$-th decoder employs the distortion measure $d_i(\cdot,\cdot): \cX \times \cXh_i\rightarrow [0,\infty)$ and wants to recover the source $x^n$ with distortion $D_i$, i.e.,
\begin{align}
d_i(x^n,\hat{X}_i^n)\leq D_i \mbox{ for $i\in\{1,2\}$}.
\end{align}
The rates of the code are defined as
\begin{align}
R_1 =&\frac{1}{n}\log M_1\\
R_2 =&\frac{1}{n}\log M_1M_2.
\end{align}

An $(n,R_1,R_2,D_1,D_2,\epsilon)$-successive refinement code is a coding scheme with block length $n$ and excess distortion probability $\epsilon$ where rates are $(R_1,R_2)$ and target distortions are $(D_1,D_2)$. Since we have two decoders, the excess distortion probability is defined by $\Pr{d_i(X^n,\hat{X}_i^n)>D_i \mbox{ for some $i$}}$. 
\begin{definition}
A rate-distortion tuple $(R_1,R_2,D_1,D_2)$ is \emph{achievable}, if there is a family of $(n, R_1^{(n)}, R_2^{(n)}, D_1, D_2,$ $\epsilon^{(n)})$-successive refinement codes where
\begin{align}
&\lim_{n\rightarrow\infty}R_i^{(n)}=R_i  \text{ for $i\in\{1,2\}$, }\\
&\lim_{n\rightarrow\infty}\epsilon^{(n)}=0.
\end{align}
\end{definition}

The achievable rate-distortion region is known:
\begin{theorem}[\cite{rimoldi1994successive}] \label{thm:rimoldi 94}
Consider a discrete memoryless source $X^n$ with law $P$. The rate-distortion tuple $(R_1,R_2,D_1,D_2)$ is achievable if and only if there is a joint law $P_{X,\Xh_1,\Xh_2}$ of random variables $(X,\Xh_1,\Xh_2)$ (where $X$ is distributed according to $P$) such that
\begin{align}
I(X;\Xh_1)\leq& R_1\\
I(X;\Xh_1,\Xh_2)\leq& R_2\\
\E{d_i(X,\hat{X}_i)}\leq& D_i \mbox{ for $i\in\{1, 2\}$}.
\end{align}
\end{theorem}

In some cases, we can achieve the optimum rates at both decoders simultaneously:
\begin{definition}\label{def:successive refinability}
For $i\in\{1,2\}$, let $R_i(P,D_i)$ denote the rate-distortion function of the source $P$ when the distortion measure is $d_i(\cdot,\cdot)$ and the distortion level is $D_i$. If $(R_1(P,D_1), R_2(P,D_2), D_1, D_2)$ is achievable, then we say the source is \emph{successively refinable at $(D_1,D_2)$}. Furthermore, if the source is successively refinable at $(D_1, D_2)$ for all non-degenerate $D_1,D_2$ (i.e., for which $R_1(P,D_1)<R_2(P,D_2)$), then we say the source is \emph{successively refinable}.
\end{definition}
A necessary and sufficient condition for successive refinability is known.
\begin{theorem}[\cite{equitz1991successive, koshelev1980hierarchical}]
A source $P$ is successively refinable at $(D_1,D_2)$ if and only if there exists a conditional distribution $P_{\Xh_1,\Xh_2|X}$ such that $X-\Xh_2-\Xh_1$ forms a Markov chain and
\begin{align}
R_i(P,D_i) &= I(X;\hat{X}_i) \\
\E{d_i(X,\hat{X}_i)}&\leq D_i
\end{align}
for $i\in\{1, 2\}$.
\end{theorem}

The condition in the theorem holds for the cases of a Gaussian source under quadratic distortion and for any discrete memoryless sources under Hamming distortion. Note that the successive refinability is not shared by all sources and distortion measures. For instance, symmetric Gaussian mixtures under quadratic distortion are not successively refinable \cite{chow1997failure}. The above results of successive refinability can be generalized to the case of $k$ decoders.

Note that we can also define successive refinability using $R(P,D_1,D_2)$ where $R(P,D_1,D_2)$ is the minimum rate $R_2$ such that $(R_1(P,D_1),R_2,D_1,D_2)$ is achievable. Using Theorem \ref{thm:rimoldi 94}, we can characterize $R(P,D_1,D_2)$,
\begin{align}
R(P,D_1,D_2) = \inf_{P_{\Xh_1,\Xh_2|X}:\substack{\fE[d_1(X,\Xh_1)]\leq D_1,\\ \fE[d_2(X,\Xh_2)]\leq D_2, \\I(X;\Xh_1)\leq R_1(P,D_1)}} I(X;\Xh_1,\Xh_2).
\end{align}
Definition \ref{def:successive refinability} implies that the source is successively refinable at $(D_1,D_2)$ if and only if $R(P,D_1,D_2) = R_2(P,D_2)$.

%%%%%%%%%%%%%%%%%%%%%%%%%%%%%%%%%%%%%%%%%%%%%%%%
%%%%%%%%%%%%%%%%           Problem Setting                            %%%%%%%%%%%%%%
%%%%%%%%%%%%%%%%%%%%%%%%%%%%%%%%%%%%%%%%%%%%%%%%
\section{Problem Setting}\label{sec:Problem Setting}
We consider the successive refinement problem with two decoders. Let $X^n = (X_1,\cdots,X_n)$ be i.i.d.\ with law $P$, where the source alphabet is $\cX$. An encoder $f^{(n)} = \left(f_1^{(n)},f_2^{(n)}\right)$ maps a source sequence to a pair of messages,
\begin{align}
&f_1^{(n)} :  \cX^n \rightarrow \{1,\cdots,M_1\}\\
&f_2^{(n)} :  \cX^n \rightarrow \{1,\cdots,M_2\}.
\end{align}

The first decoder receives only the output of $f_1^{(n)}(X^n)$, and therefore we say that its rate is $R_1 =  (1/n)\log M_1$. The second decoder receives the output of both functions, so its rate is $R_2 = (1/n)\log M_1M_2$.

Decoder 1 employs a decoder $g_1^{(n)} : \{1,\cdots,M_1\}\rightarrow \cXh_1^n$ and decoder 2 employs a decoder $g_2^{(n)} : \{1,\cdots,M_1\}\times\{1,\cdots,M_2\}\rightarrow \cXh_2^n$, where $\cXh_1$ and $\cXh_2$ are the reconstruction alphabets for each decoder. Decoder $i$ has its own distortion measure $d_i: \cX\times \cXh_i\rightarrow [0,\infty)$ with a target distortion $D_i$. Both $d_1$ and $d_2$ are symbol by symbol distortion measures which induce block distortion measures by
\begin{align}
d_i(x^n,\hat{x}_i^n) = \frac{1}{n}\sum_{j=1}^n d_i(x_j,\hat{x}_{i,j})
\end{align}
for all $i\in\{1,2\}$, $x^n\in\cX^n$, $\hat{x}_1^n\in\cXh_1^n$ and $\hat{x}_2^n\in\cXh_2^n$. The setting is described in Figure \ref{fig:Successive Refinement}.
\tikzstyle{format} = [thin]
\tikzstyle{medium} = [rectangle, draw, thin, fill=blue!20, minimum height=2.5em, minimum width = 4em]
\begin{figure}[ht]
\centering
\begin{tikzpicture}[node distance=2.8cm, auto,>=latex', thick]
    \node [format] (src) {$X^n$};
    \node [medium, right of=src, node distance = 2cm](enc){Enc};
    \node [medium, right of=enc, node distance = 3cm](dec1){Dec 1};
    \node [medium, below of=dec1, node distance = 1.5cm](dec2){Dec 2};
    \node [format, right of=dec1, node distance = 2cm](rec1){$\Xh_1^n$};
    \node [format, right of=dec2, node distance = 2cm](rec2){$\Xh_2^n$};

    \draw [->] (src) -- node {}(enc);
    \draw [->] (enc) -- node {$m_1$}(dec1);
   \draw  [->] (enc.east) -- ++(.7,0)  -- ++(0,-1.3)-- ++(.87,0);
    \draw [->] (enc) |- node [above right] {$m_2$}(dec2);
    \draw [->] (dec1) -- node{} (rec1);
    \draw [->] (dec2) -- node{} (rec2);
\end{tikzpicture}
\caption{Successive Refinement}\label{fig:Successive Refinement}
\end{figure}
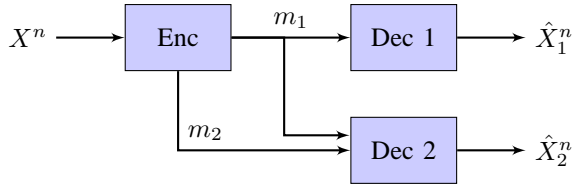

\begin{definition}
We say that $(n,M_1,M_2,D_1,D_2,\epsilon_1,\epsilon_2)$ is \emph{achievable} if there exists an encoder-decoder pair that satisfies
\begin{align}
\Pr{d_1(X^n,g_1^{(n)}(f_1^{(n)}(X^n))) > D_1} \leq& \epsilon_1\\
\Pr{d_2(X^n,g_2^{(n)}(f_1^{(n)}(X^n),f_2^{(n)}(X^n))) > D_2} \leq& \epsilon_2,
\end{align}
and such a code is called a $(n,M_1,M_2,D_1,D_2,\epsilon_1,\epsilon_2)$-code.
\end{definition}

Note that we consider the \emph{two} error events separately, unlike in the definition of a successive refinement code in Section \ref{subsec:Successive Refinement}. Our goal is to characterize the achievable $(n,M_1,M_2,D_1,D_2,\epsilon_1,\epsilon_2)$ region in general. Motivated by successive refinability, we define strong successive refinability as follows.
\begin{definition}
The source is \emph{strongly successively refinable} at $(D_1,D_2,\epsilon_1,\epsilon_2$) if $(n,M_1, M_2, D_1,D_2, \epsilon_1, \epsilon_2)$ is achievable for some $M_1,M_2$ satisfying 
\begin{align}
&\frac{1}{n}\log M_1 = R_1(P,D_1) + \sqrt{\frac{V_1(P,D_1)}{n}}Q^{-1}(\epsilon_1) + o\left(\frac{1}{\sqrt{n}}\right)\\
&\frac{1}{n}\log M_1M_2 = R_2(P,D_2) + \sqrt{\frac{V_2(P,D_2)}{n}}Q^{-1}(\epsilon_2) + o\left(\frac{1}{\sqrt{n}}\right)
\end{align}
where $R_i(P,D_i)$ and $V_i(P,D_i)$ are the point-to-point rate-distortion function and the source dispersion for the $i$-th decoder. Furthermore, if the source is strongly successively refinable at $(D_1,D_2,\epsilon_1,\epsilon_2)$ for all non-degenerate $D_1,D_2,\epsilon_1,\epsilon_2$ (i.e., $R_{P,D_1,\epsilon}(n) < R_{P,D_2,\epsilon}(n)$), then we say the source is \emph{strongly successively refinable}.
\end{definition}

While standard successive refinability implies that the successive refinement structure does not cause any loss in the compression rate (asymptotically), strong successive refinability implies that we also do not lose from the dispersion point of view.

Note that in order to verify that a source is strongly successively refinable, it is sufficient to find an achievability scheme since the converse will follow from the converse in point-to-point source coding.

%%%%%%%%%%%%%%%%%%%%%%%%%%%%%%%%%%%%%%%%%%%%%%%%
%%%%%%%%%%%%%%%%           Main Results                            %%%%%%%%%%%%%%
%%%%%%%%%%%%%%%%%%%%%%%%%%%%%%%%%%%%%%%%%%%%%%%%
\section{Main Results}\label{sec:Main Results}
Our results in this section pertain to discrete memoryless sources under general distortion, as well as Gaussian sources under quadratic distortion. The results are given here, with proofs in Section \ref{sec:Proof}.

%%%%%%%%%%%%%%%%       Discrete Memoryless Source
%%%%%%%%%%%%%%%%%%%%%%%%
\subsection{Discrete Memoryless Source}\label{subsec:Discrete Memoryless Source}
Let $X^n$ be i.i.d.\ with distribution $P$ and the distortion measures be $d_1:\cX \times\cXh_1 \rightarrow[0,\infty)$ and $d_2:\cX \times\cXh_2 \rightarrow[0,\infty)$. We assume that the alphabets $\cX$, $\cXh_1$ and $\cXh_2$ are finite, and therefore distortion measures $d_1$ and $d_2$ are bounded by some constant $d_M$. We further assume that $P(x)>0$ for all $x\in\cX$ since one can remove the source symbol from $\cX$ that has zero probability. Then, the following theorem provides the achievable rates including the second order term:

\begin{theorem}[Achievability for Discrete Memoryless Source]\label{thm:AchievabilityDMS}
Assume that both $R_1(P,D_1)$ and $R(P,D_1,D_2)$ are continuously twice differentiable with respect to $D_1, D_2$, and the elements of $P$ in some neighborhood of $(P,D_1,D_2)$. Then, there exists an $(n,M_1,M_2,D_1,D_2,\epsilon_1,\epsilon_2)$-code such that
\begin{align}
&\frac{1}{n}\log M_1 = R_1(P,D_1) + \sqrt{\frac{V_1(P,D_1)}{n}}Q^{-1}(\epsilon_1) + O\left(\frac{\log n}{n}\right)\\
&\frac{1}{n}\log M_1M_2= R(P,D_1,D_2) + \sqrt{\frac{V(P,D_1,D_2)}{n}}Q^{-1}(\epsilon_2) + O\left(\frac{\log n}{n}\right)
\end{align}
where
\begin{align}
V_1(P,D_1) \triangleq& \Var{R'_1(X,D_1)}\\
=& \sum_{x\in\cX} P(x)(R_1'(x,D_1))^2 - \left[\sum_{x\in\cX} P(x)R_1'(x,D_1)\right]^2\\
V(P,D_1,D_2) \triangleq& \Var{R'(X,D_1,D_2)}\\
=& \sum_{x\in\cX} P(x)(R'(x,D_1,D_2))^2 - \left[\sum_{x\in\cX} P(x)R'(x,D_1,D_2)\right]^2.
\end{align}
Similarly to Theorem \ref{thm:IngberDispersion}, $R_1'(x,D_1)$ is the derivative of $R_1(P,D_1)$ with respect to the probability $P(x)$ and $R'(x,D_1,D_2)$ is the derivative\footnote{Similar to the definition of $R'(x,D)$, we can define $R'(x,D_1,D_2)$ using an extension. Then, $R'(x,D_1,D_2)$ and $V(P,D_1,D_2)$ are well and uniquely-defined as well, where details are given in Appendix \ref{app:Derivative of Rate-Distortion Function}.} of $R(P,D_1,D_2)$ with respect to the probability $P(x)$:
\begin{align}
R_1'(x,D_1)\triangleq& \left[\frac{\partial R_1(Q,D_1)}{\partial Q(x)}\right]_{Q=P} \\
R'(x,D_1,D_2)\triangleq &\left[\frac{\partial R(Q,D_1,D_2)}{\partial Q(x)}\right]_{Q=P}.
\end{align}
\end{theorem}

By applying the above theorem to the special case where $R(\tilde{P},D_1,D_2) = R_2(\tilde{P},D_2)$ for all $\tilde{P}$ in some neighborhood of $P$, we get the following corollary.
\begin{corollary}\label{cor:Achievability Strong Refinability}
Suppose $R_i(P,D_i)$ is continuously twice differentiable with respect to $D_i$ and the elements of $P$ in some neighborhood of $(P,D_i)$ for $i\in\{1,2\}$. If all sources $\tilde{P}$ in some neighborhood of $P$ are successively refinable at $D_1,D_2$, then there exists an $(n,M_1,M_2,D_1,D_2,\epsilon_1,\epsilon_2)$ code such that
\begin{align}
&\frac{1}{n}\log M_1= R_1(P,D_1) + \sqrt{\frac{V_1(P,D_1)}{n}}Q^{-1}(\epsilon_1) + O\left(\frac{\log n}{n}\right)\\
&\frac{1}{n}\log M_1M_2= R_2(P,D_2) + \sqrt{\frac{V_2(P,D_2)}{n}}Q^{-1}(\epsilon_2) + O\left(\frac{\log n}{n}\right),
\end{align}
i.e., the source $P$ is strongly successively refinable at $(D_1,D_2,\epsilon_1,\epsilon_2)$. 
\end{corollary}
The corollary is because $R(\tilde{P},D_1,D_2) = R_2(\tilde{P},D_2)$ for all $\tilde{P}$ in some neighborhood of $P$ implies that their derivatives at $(P,D_1,D_2)$ coincide, i.e., 
\begin{align}
\left[\frac{\partial R_2(Q,D_2)}{\partial Q(x)}\right]_{Q=P}=\left[\frac{\partial R(Q,D_1,D_2)}{\partial Q(x)}\right]_{Q=P}.
\end{align}
Since the source dispersion is the variance of the derivatives, we have $V(P,D_1,D_2) = V_2(P,D_2)$.

\begin{remark}
Any discrete memoryless source under Hamming distortion measure is successively refinable. Therefore, Corollary \ref{cor:Achievability Strong Refinability} implies that any discrete memoryless source under Hamming distortion is also \emph{strongly} successively refinable provided $R(P,D)$ is appropriately differentiable. Note that the size of the set $\{D:R(P,D)$ is not differentiable$\}$ is at most $\card{\cX}$ \cite{erokhin1958varepsilon}.
\end{remark}

%%%%%%%%%%%%%%%%       Gaussian Memoryless Source
%%%%%%%%%%%%%%%%%%%%%%%%
\subsection{Gaussian Memoryless Source}\label{subsec:Gaussian Memoryless Source}
Let $X^n$ be an i.i.d.\ $\cN(0,\sigma^2)$ source, and suppose the distortion measure is quadratic (at both decoders).  

\begin{theorem}[Achievability for Gaussian Memoryless Source]\label{thm:AchievabilityGMS}
The memoryless Gaussian source under quadratic distortion is strongly successively refinable, i.e., for $\sigma^2>D_1>D_2$, there exists an $(n,M_1,M_2,D_1,D_2,\epsilon_1,\epsilon_2)$ code such that
\begin{align}
&\frac{1}{n}\log M_1=  \frac{1}{2}\log \frac{\sigma^2}{D_1} + \sqrt{\frac{1}{2n}}Q^{-1}(\epsilon_1) + O\left(\frac{\log n}{n}\right)\\
&\frac{1}{n}\log M_1M_2 =\frac{1}{2}\log \frac{\sigma^2}{D_2}+ \sqrt{\frac{1}{2n}}Q^{-1}(\epsilon_2) + O\left(\frac{\log n}{n}\right).
\end{align}
\end{theorem}

%%%%%%%%%%%%%%%%%%%%%%%%%%%%%%%%%%%%%%%%%%%%%%%%
%%%%%%%%%%%%%%%%           Proof                            %%%%%%%%%%%%%%%%%%%
%%%%%%%%%%%%%%%%%%%%%%%%%%%%%%%%%%%%%%%%%%%%%%%%
\section{proof}\label{sec:Proof}

%%%%%%%%%%%%%%%%       Method of Types
%%%%%%%%%%%%%%%%%%%%%%%%
\subsection{Method of Types}\label{subsec:Method of Types}
Our proofs for finite alphabet sources rely heavily on the method of types \cite{csiszar2011information}. In this section, we briefly review its notation and results that we use.  Without loss of generality we assume $\cX = \{1,2,\ldots,r_x\}$. For any sequence $x^n\in\cX^n$, let $N(a|x^n)$ be the number of symbol $a\in\cX$ in the sequence $x^n$. Let the \emph{type} of a sequence $x^n$ be an $r_x$ dimensional vector $P_{x^n}=\left({N(1|x^n)}/{n}, {N(2|x^n)}/{n},\ldots,{N(r_x|x^n)}/{n}\right)$. Then, denote $\cP_n(\cX)$ be the set of all types on $\cX^n$, i.e., $\cP_n(\cX) = \{P_{x^n}\suchthat x^n\in\cX^n\}$. The size of the set $\cP_n(\cX)$ is at most polynomial in $n$, more precisely,
\begin{align}
\card{\cP_n(\cX)}\leq (n+1)^{r_x}. \label{eq:number of types}
\end{align}

For given type $P$, define \emph{type class} of $P$ by
\begin{align}
\cT_P = \{x^n\in\cX^n\suchthat P_{x^n} = P\}.
\end{align}
We can also define type class  $\cT_{x^n} = \{\tilde{x}^n\in\cX^n\suchthat P_{\tilde{x}^n} = P_{x^n}\}$ using a sequence $x^n\in\cX^n$. We can bound the size of type class.
\begin{align}
\frac{1}{(n+1)^{r_x}}\exp\left(nH(P)\right)\leq \card{\cT_P}\leq \exp\left(nH(P)\right) \label{eq:size of type classes}
\end{align}
where $H(P)$ denote an entropy of random variable with law $P$.

We further consider the conditional types. Let $\cY$ be a set of alphabet where we also assume $\cY=\{1,2,\ldots,r_y\}$ to be finite. Consider a stochastic kernel $W:\cX\rightarrow \cY$. We say that $y^n\in\cY^n$ has \emph{conditional type} $W$ given $x^n\in\cX^n$ if
\begin{align}
N(a,b|x^n,y^n)= N(a|x^n) W(b|a).
\end{align}
Then, we can define \emph{conditional type class} of $W$ given $x^n\in\cX^n$ by
\begin{align}
\cT_W(x^n) = \{y^n\in\cY^n\suchthat\mbox{$y^n$ has conditional type $W$ given $x^n$}\}.
\end{align}

We can also bound a size of conditional type class. For sequence $x^n\in\cX^n$ with type $P$, and for conditional type $W$, we have
\begin{align}
\frac{1}{(n+1)^{r_x r_y}}\exp\left(nH(P|W)\right)\leq \card{\cT_W(x^n)}\leq \exp\left(nH(P|W)\right).\label{eq:size of conditional type classes}
\end{align}
$H(P|W)$ denotes a conditional entropy of $U$ given $V$ where $(U,V)$ are random variables with a joint law $P\times W$.

%%%%%%%%%%%%%%%%       Proof of Achievability
%%%%%%%%%%%%%%%%%%%%%%%%
\subsection{Proof of Theorem~\ref{thm:AchievabilityDMS}}\label{subsec:Proof of Achievability DMS}
A key tool used in the proof is a refined version of the type covering lemma \cite{csiszar2011information}. We say a set $B$ is $D$-covering a set $A$ if for all $a\in A$, there exists an element $b\in B$ such that $d(a,b)\leq D$. In the successive refinement setting, we need to cover a set in a successive manner. 
\begin{definition}
Let $d_1: \cA \times \cB \rightarrow [0,\infty)$ and $d_2:\cA \times \cC \rightarrow [0,\infty)$ be distortion measures. Consider sets $A\subset \cA$, $B\subset \cB$ and $C_b\subset\cC$ for all $b\in B$. We say $B$ and $\{C_b\}_{b\in B}$ \emph{successively $(D_1,D_2)$-cover} a set $A$, if for all $a\in A$, there exist $b\in B$ and $c\in C_b$ such that $d_1(a,b) \leq D_1$ and $d_2(a,c)\leq D_2$.
\end{definition}

The following lemma provides an upper bound of minimum size of sets that successively $(D_1,D_2)$-cover a type class $\cT_P$.
\begin{lemma}[Refined Covering Lemma]\label{lem:GeneralizedRefinedCovering}
For fixed $n$, let $P\in \cP_n(\cX)$ be a type on $\cX$ where $P(x)>{3}/{n}$ for all $x\in\cX$. Suppose $\norm{\nabla R(P,D_1,D_2)}$ is bounded in some neighborhood of $(D_1,D_2)$ where
\begin{align}
\nabla R(P,D_1,D_2)=\left(\frac{\partial}{\partial D_1} R(P,D_1,D_2),\frac{\partial}{\partial D_2} R(P,D_1,D_2)\right).
\end{align}
Then for $D_1,D_2\in (0,d_M)$, there exist sets $B_1\subset \cXh_1^n$ and $B_2(\hat{x}_1^n)\subset \cXh ^n_2$ for each $\hat{x}_1^n\in B_1$ where $B_1$ and $\{B_2(\hat{x}_1^n)\}_{\hat{x}_1^n\in B_1}$ successively $(D_1,D_2)$-cover $\cT_P$ with following properties:
\begin{itemize}
\item The size of $B_1$ is upper bounded:
\begin{align}
\frac{1}{n}\log \card{B_1} \leq& R_1(P,D_1)+k_1\frac{\log n}{n}.\label{eq:bound BPD1}
\end{align}
\item For all $\hat{x}_1^n \in B_1$, the size of $B_2(\hat{x}_1^n)$ is also bounded:
\begin{align}
\frac{1}{n}\log\left(\card{B_1}\cdot\card{B_2(\hat{x}_1^n)} \right)\leq& R(P,D_1,D_2)+k_2\frac{\log n}{n},\label{eq:bound BPD1D2}
\end{align}
where $k_1$ and $k_2$ are universal constants, i.e., do not depend on the distribution $P$ or $n$.
\end{itemize}
\end{lemma}
The proof of Lemma \ref{lem:GeneralizedRefinedCovering} is given in Appendix \ref{app:Proof of Generalized Refined Covering Lemma for Successive Refinement}. The following corollary provides a successive refinement scheme using $B_1$ and $\{B_2(\hat{x}_1^n)\}_{\hat{x}_1^n\in B_1}$ from Lemma \ref{lem:GeneralizedRefinedCovering}.
\begin{corollary}\label{cor:message splitting}
For length of sequence $n$ and type $Q\in \cP_n(\cX)$, let $\tilde{R}$ satisfy $\tilde{R} \geq R_1(Q,D)+k_1{\log n}/{n}$. Then, there exists a coding scheme for $\cT_Q$ such that
\begin{itemize}
\item Encoding functions are $f_{Q,1}: \cT_Q \rightarrow \{1,\ldots,M_{Q,1}\}$ and $f_{Q,2} : \cT_Q \rightarrow \{1,\ldots,M_{Q,2}\}$.
\item Decoder 1 and Decoder 2 employ
\begin{align}
&g_{Q,1}: \{1,\ldots,M_{Q,1}\} \rightarrow \cXh_1^n\\
&g_{Q,2}:\{1,\ldots,M_{Q,1}\}\times\{1,\ldots,M_{Q,2}\}\rightarrow \cXh_2^n
\end{align} 
respectively.
\item For all $x^n\in \cT_Q$, encoding and decoding functions satisfy
\begin{align}
d_1\left(x^n,g_{Q,1}( f_{Q,1}(x^n))\right)\leq& D_1 \label{eq:encoding for type q1}\\
d_2\left(x^n,g_{Q,2}(  f_{Q,1}(x^n), f_{Q,2}(x^n))\right)\leq& D_2.\label{eq:encoding for type q2}
\end{align}
\item The number of messages are bounded:
\begin{align}
\tilde{R}\leq \frac{1}{n} \log M_{Q,1} \leq& \tilde{R}+\frac{\log n}{n}\label{eq: rate bound for coding for typeq1}\\
\frac{1}{n} \log M_{Q,1}M_{Q,2} \leq& R(Q,D_1,D_2)+(k_2+1)\frac{\log n}{n}.\label{eq: rate bound for coding for typeq2}
\end{align}
\end{itemize}
\end{corollary}
The proof of Corollary \ref{cor:message splitting} is given in Appendix \ref{app:Proof of Corollary}.

Let us now describe the achievability scheme. Similar to the idea from \cite{tuncel2003error}, we will consider the four cases according to the type $Q$ of the input sequence $x^n$. For each case, the encoding will be done in a different manner. Before specifying four cases, we need to define $\Delta R_1$ and $\Delta R_2$. Let $\Delta R_1$ be the infimal value such that the probability of $\{R_1(P_{X^n},D_1) > R_1(P,D_1)+\Delta R_1\}$ is smaller than $\epsilon_1$, and $\Delta R_2$ be the infimal value such that the probability of $\{R(P_{X^n},D_1,D_2) > R(P,D_1,D_2)+\Delta R_2\}$ is smaller than $\epsilon_2$. Recall that $P_{X^n}$ denotes the type of $X^n$. The error occurs at decoder 1 if and only if $R_1(P_{X^n},D_1) > R_1(P,D_1)+\Delta R_1$, and therefore probability of error at decoder 1 is less than $\epsilon_1$. Similarly, the error occurs at decoder 2 if and only if $R(P_{X^n},D_1,D_2)>R(P,D_1,D_2)+\Delta R_2$, and therefore probability of error at decoder 2 is less than $\epsilon_2$. The following lemma bounds $\Delta R_1$ and $\Delta R_2$.
\begin{lemma}\label{lem:BoundDeltaR}
\begin{align}
\Delta R_1 =& \sqrt{\frac{V_1(P,D_1)}{n}}Q^{-1}(\epsilon_1)+O\left(\frac{\log n}{n}\right)\\
\Delta R_2 =& \sqrt{\frac{V(P,D_1,D_2)}{n}}Q^{-1}(\epsilon_2)+O\left(\frac{\log n}{n}\right).
\end{align}
\end{lemma}
The proof follows directly from \cite[Lemma 3]{DBLP:journals/corr/abs-1109-6310}.

We are ready to define four cases based on the type of the source sequence as well as corresponding encoding schemes.
\begin{enumerate}
\item  $Q\in A^{(0,0)} \triangleq \{Q\in \cP_n(\cX): R_1(Q,D_1)-R_1(P,D_1) \leq \Delta R_1, R(Q,D_1,D_2)-R(P,D_1,D_2) \leq \Delta R_2\}$.\\
In this case, both decoders decode successfully. Since $R(Q,D_1)\leq R(P,D_1) + \Delta R_1$, by Corollary \ref{cor:message splitting}, there exist encoding and decoding functions $f_{Q,1}, f_{Q,2}, g_{Q,1}, g_{Q,2}$ such that
\begin{align}
d_1\left(x^n,g_{Q,1}( f_{Q,1}(x^n))\right)\leq& D_1 \\
d_2\left(x^n,g_{Q,2}(  f_{Q,1}(x^n), f_{Q,2}(x^n))\right)\leq& D_2
\end{align}
for all $x^n\in \cT_Q$ and 
\begin{align}
R_1(P,D_1)+\Delta R_1+k_1\frac{\log n}{n}\leq\frac{1}{n}\log M_{Q,1}^{(0,0)} \leq& R_1(P,D_1)+\Delta R_1+(k_1+1)\frac{\log n}{n}\label{eq:PnotQ}\\
\frac{1}{n}\log M_{Q,1}^{(0,0)} M_{Q,2}^{(0,0)}\leq& R(Q,D_1,D_2)+(k_2+1)\frac{\log n}{n}.
\end{align}
We emphasize that we have $R_1(P,D_1)$ instead of $R_1(Q,D_1)$ in \eqref{eq:PnotQ}. This is because we need to aggregate the codewords at the end of the proof. More precisely, we have to fix the number of codewords for decoder 1 in order to bound the number of codewords only for decoder 2.

\item $Q\in A^{(0,1)} \triangleq \{Q\in \cP_n(\cX): R_1(Q,D_1)-R_1(P,D_1) \leq \Delta R_1, R(Q,D_1,D_2)-R(P,D_1,D_2) > \Delta R_2\}$.\\
For those $Q$, the encoder only $D_1$ covers $\cT_Q$. Thus, decoder 1 will decode successfully and decoder 2 will declare an error. In this case, we do not need a message for decoder 2 and we can think of $M_2^{(0,1)} = 1$. For decoder 1,  by Theorem \ref{thm:IngberDispersion}, we can find encoding and decoding functions $f^{(0,1)}: \cX^n\rightarrow \{1,\ldots, M_1^{(0,1)}\}$ and $g^{(0,1)}: \{1,\ldots,M_1^{(0,1)}\}\rightarrow \cXh_1^n$ such that 
\begin{align}
d_1(x^n,g^{(0,1)}(f^{(0,1)}(x^n)))\leq D_1
\end{align}
for all $Q\in A^{(0,1)}$ and $x^n\in \cT_Q$ where
\begin{align}
\frac{1}{n}\log M_1^{(0,1)} = R_1(P,D_1)+\Delta R_1 + O\left(\frac{\log n}{n}\right).
\end{align}

\item $Q\in A^{(1,0)} \triangleq \{Q\in \cP_n(\cX): R_1(Q,D_1)-R_1(P,D_1) > \Delta R_1, R(Q,D_1,D_2)-R(P,D_1,D_2) \leq \Delta R_2\}$.\\
In this case, the encoder only $D_2$ covers $\cT_Q$. Thus, decoder 2 will decode successfully and decoder 1 will declare an error. In this case, we do not need a message for decoder 1. However, because of the structure of successive refinement code, we need to reformulate the point-to-point code for the second decoder into the form of successive refinement code. More precisely, we can find functions $\tilde{f}_Q^{(1,0)} : \cX^n \rightarrow \{1,\ldots,\tilde{M}_{Q,2}^{(1,0)}\}$ and $\tilde{g}_{Q}^{(1,0)} :\{1,\ldots,\tilde{M}_{Q,2}^{(1,0)}\} \rightarrow \Xh_2^n$ such that
\begin{align}
d_2(x^n,\tilde{g}_Q^{(1,0)}(\tilde{f}_Q^{(1,0)}(x^n)))\leq D_2
\end{align}
for all $x^n\in \cT_Q$ where
\begin{align}
\frac{1}{n}\log \tilde{M}_{Q,2}^{(1,0)} \leq R_2(Q,D_2)+k_1\frac{\log n}{n}.
\end{align}

Let $M_{Q,1}^{(1,0)}$ and $M_{Q,2}^{(1,0)}$ be 
\begin{align}
R_1(P,D_1)+\Delta R_1+k_1\frac{\log n}{n}\leq \frac{1}{n}\log M^{(1,0)}_{Q,1} \leq& R_1(P,D_1)+\Delta R_1+(k_1+1)\frac{\log n}{n}\\
\frac{1}{n}\log M^{(1,0)}_{Q,1}M^{(1,0)}_{Q,2}\leq& \frac{1}{n}\log \tilde{M}_{Q,2}^{(1,0)}+\frac{\log n}{n}.
\end{align}
For simplicity, we neglect the fact that the number of messages are integers since it will increase the rate by at most ${\log n}/{n}$ bits/symbol. Let $h$ be a one to one mapping from $\{1,\ldots, M_{Q,1}^{(1,0)}\}\times \{1,\ldots, M_{Q,2}^{(1,0)}\}$ to $\{1,\ldots, \tilde{M}_{Q,2}^{(1,0)}\}$. Then, we can define encoding and decoding functions $f_{Q,1}^{(1,0)}:\cX^n\rightarrow \{1,\ldots, M_{Q,1}^{(1,0)}\}$, $f_{Q,2}^{(1,0)}:\cX^n \rightarrow \{1,\ldots, M_{Q,2}^{(1,0)}\}$, and $g_{Q,2}^{(1,0)} :\{1,\ldots, M_{Q,1}^{(1,0)}\}\times\{1,\ldots, M_{Q,2}^{(1,0)}\} \rightarrow \cXh_2^n$ where 
\begin{align}
\left(f_{Q,1}^{(1,0)}(x^n) ,f_{Q,1}^{(1,0)}(x^n) \right) =& h^{-1} \left(\tilde{f}_Q^{(1,0)}(x^n)\right)\\
g_{Q,2}^{(1,0)}(m_1,m_2) =& \tilde{g}_Q^{(1,0)}(h(m_1,m_2)).
\end{align}

Note that the first message is useless for decoder 1, but we do not care since it will declare an error anyway. On the other hand, decoder 2 will decode both $m_1$ and $m_2$ successfully.

\item $Q\in A^{(1,1)} \triangleq \{Q\in \cP_n(\cX): R_1(Q,D_1)-R_1(P,D_1) > \Delta R_1, R(Q,D_1,D_2)-R(P,D_1,D_2) > \Delta R_2\}$.\\
The encoder sends nothing and the both decoder will declare errors. We can assume $M_1^{(1,1)} = M_2^{(1,1)} = 1$.
\end{enumerate}

Finally, we merge all encoding functions together. Given source sequence $x^n$, the encoder describes a type of sequence as a part of the first message using $\card{\cX} \log (n+1)$ bits. This affects at most $O({\log n}/{n})$ bits/symbol in rates. Based on the type of sequence, it employs an encoding function accordingly, as described above. Since the decoder also knows the type of the sequence, it can employ the corresponding decoding function. Since all $M_{Q,1}^{(0,0)}, M_1^{(0,1)}, M_{Q,1}^{(1,0)}$ have the same upper bound, we can bound $M_1$:
\begin{align}
\frac{1}{n}\log M_1 \leq&  R_1(P,D_1) + \Delta R_1+ (k_1+1)\frac{\log n}{n}+\card{\cX} \frac{\log (n+1)}{n}\\
\leq& R_1(P,D_1)+\sqrt{\frac{V_1(P,D_1)}{n}}Q^{-1}(\epsilon_1) + O\left(\frac{\log n}{n}\right).
\end{align}
Similarly, we can show that 
\begin{align}
&\frac{1}{n}\log M_1M_2\leq R(P,D_1,D_2)+\sqrt{\frac{V(P,D_1,D_2)}{n}}Q^{-1}(\epsilon_2) + O\left(\frac{\log n}{n}\right).
\end{align}
This concludes the proof.  

%%%%%%%%%%%%%%%%       Proof of Achievabiliy
%%%%%%%%%%%%%%%%%%%%%%%%
\subsection{Proof of Theorem~\ref{thm:AchievabilityGMS}}\label{subsec:Proof of Achievability GMS} 
Instead of type covering arguments that we used in the previous section, we use the result of sphere covering for Gaussian sources.
\begin{theorem}[\cite{rogers1963covering}]\label{thm:sphere covering}
There is an absolute constant $k_s$ such that, if $R>1$ and $n\geq 9$, any $n$-dimensional spheres of radius $R$ can be covered by less than $k_s n^{5/2} R^n$ spheres of radius 1.
\end{theorem}
For simplicity, we refer to the sphere of radius $r$ by $r$-ball and denote by $\cB(x^n,r) = \{\tilde{x}^n : d(x^n,\tilde{x}^n)\leq r^2\}$, the set of points in the sphere centered at $x^n$ with radius $r$. The above theorem immediately implies the following corollary.
\begin{corollary}\label{cor:sphere covering}
For $n\geq 9$ and $R_1>R_2>0$, we can find a set $\cC\subset \fR^n$ of size $M$ that satisfies:
\begin{itemize}
\item For all $x^n\in \cB(0,R_1)$, there is an element $\hat{x}^n\in \cC$ such that $x^n \in \cB(\xh^n, R_2)$.
\item The size of the set $M$ is upper bounded by 
\begin{align}\frac{1}{n}\log M \leq \frac{1}{2} \log \frac{R_1}{R_2} + \frac{5}{2}\frac{\log n}{n}+O\left(\frac{1}{n}\right).
\end{align}
\end{itemize}
\end{corollary}

Let $r_1$ and $r_2$ be radius of the balls such that $\Pr{X_1^2+\cdots+X_n^2 > r_1^2}= \epsilon_1$ and $\Pr{X_1^2+\cdots+X_n^2 > r_2^2} = \epsilon_2$. First, consider the case $\epsilon_1<\epsilon_2$. It is clear that $Q^{-1}(\epsilon_1)>Q^{-1}(\epsilon_2)$ and $r_1>r_2$.  We can further divide this case into the following three cases,
\begin{enumerate}
\item $X^n \in \cB({\bf 0}, r_2)$, i.e., $X_1^2+\cdots+X_n^2 \leq r_2^2$. In this case, we design a code such that both decoders can decode successfully.\\
Let $\cC_1^{(0,0)} \subset \fR^n$ be the set that satisfies:
\begin{itemize}
\item $\card{\cC_1^{(0,0)}} = M_1^{(0,0)}$
\item $\cB({\bf 0},r_2) \subset \bigcup_{\hat{x}^n\in \cC_1^{(0,0)}}\cB(\hat{x}^n, \sqrt{nD_1})$
\item $\frac{1}{n}\log M_1^{(0,0)}\leq \frac{1}{2}\log \frac{\sigma^2}{D_1}+\sqrt{\frac{1}{2n}} Q^{-1}(\epsilon_2) +O\left(\frac{\log n}{n}\right)$
\end{itemize}
which implies that there are $M_1^{(0,0)}$ number of $\sqrt{nD_1}$-balls that covers the $r_2$-ball. Upper bound on $M^{(0,0)}_1$ can be found similarly to the proof of Theorem \ref{thm:IngberDispersionGaussian}. Since $Q^{-1}(\epsilon_1)>Q^{-1}(\epsilon_2)$, it is clear that 
\begin{align}
\frac{1}{n}\log M_1^{(0,0)}\leq \frac{1}{2}\log \frac{\sigma^2}{D_1}+\sqrt{\frac{1}{2n}} Q^{-1}(\epsilon_1) +O\left(\frac{\log n}{n}\right).
\end{align}

Similarly, we can cover a $\sqrt{nD_1}$-ball with $M_2^{(0,0)}$ number of $\sqrt{nD_2}$-balls. In other words, there exists $\cC_2^{(0,0)} \subset \fR^n$ that satisfies:
\begin{itemize}
\item $\card{\cC_2^{(0,0)}} = M_2^{(0,0)}$
\item $\cB({\bf 0},\sqrt{nD_1}) \subset \bigcup_{\hat{x}^n\in \cC_2^{(0,0)}}\cB(\hat{x}^n, \sqrt{nD_2})$
\item $\frac{1}{n}\log M_2^{(0,0)}\leq \frac{1}{2}\log\frac{D_1}{D_2} +O\left(\frac{\log n}{n}\right)$
\end{itemize} 
where upper bound on $M_2^{(0,0)}$ is because of Corollary \ref{cor:sphere covering}.

Thus, if $x^n\in \cB({\bf 0},r_2)$, then we can find $\hat{x}_1^n \in \cC_1^{(0,0)}$  such that $x^n\in \cB(\hat{x}_1^n,\sqrt{nD_1})$ which implies $({1}/{n})\norm{x^n-\hat{x}_1^n}_2^2\leq D_1$. Furthermore, since $x^n-\hat{x}_1^n\in \cB({\bf 0}, \sqrt{nD_1})$, we can find $\tilde{x}^n\in \cC_2^{(0,0)}$ such that $x^n-\hat{x}_1^n \in \cB(\tilde{x}^n,\sqrt{nD_2})$ which implies $({1}/{n})\norm{x^n-\hat{x}_1^n-\tilde{x}^n}_2^2\leq D_2$. Finally, we can take $\hat{x}_2^n = \hat{x}_1^n + \tilde{x}^n$, and we get $({1}/{n})\norm{x^n-\hat{x}_2^n}_2^2 \leq D_2$.

\item $X^n \in \cB({\bf 0}, r_1)$ but $X^n \notin \cB({\bf 0}, r_2)$, i.e., $r_2^2<X_1^2+\cdots+X_n^2 \leq r_1^2$.\\
We will only send a message to decoder 1, and decoder 2 will declare an error. We can cover $r_1$-ball with $M_1^{(0,1)}$ number of $\sqrt{nD_1}$-balls where 
\begin{align}
\frac{1}{n}\log M_1^{(0,1)}\leq \frac{1}{2}\log \frac{\sigma^2}{D_1}+\sqrt{\frac{1}{2n}} Q^{-1}(\epsilon_1) +O\left(\frac{\log n}{n}\right).
\end{align}
Therefore, there exists $\cC_1^{(0,1)}$ that satisfies:
\begin{itemize}
\item $\card{\cC_1^{(0,1)}} = M_1^{(0,1)}$
\item $\cB({\bf 0},r_2) \subset \bigcup_{\hat{x}^n\in \cC_1^{(0,1)}}\cB(\hat{x}^n, \sqrt{nD_1})$
\item $\frac{1}{n}\log M_1^{(0,1)}\leq \frac{1}{2}\log \frac{\sigma^2}{D_1}+\sqrt{\frac{1}{2n}} Q^{-1}(\epsilon_1) +O\left(\frac{\log n}{n}\right).$
\end{itemize}
 We can think $M_2^{(0,1)}$ to be one.

\item $X^n \notin \cB({\bf 0}, r_1)$ and $X^n \notin \cB({\bf 0}, r_2)$, i.e., $r_1^2<X_1^2+\cdots+X_n^2 $.\\
The encoder does not send any messages and both decoder will declare an error. We can think both $M_1^{(0,2)}$ and $M_2^{(0,2)}$ to be one.
\end{enumerate}
Finally, we employ the codebook $\cC_1 = \cC_1^{(0,0)}\cup \cC_1^{(0,1)}\cup\cC_1^{(0,2)}$ and the same for $\cC_2$ where $\card{\cC_1} = M_1$ and $\card{\cC_2} = M_2$. Then, we can see that 
\begin{align}
\frac{1}{n} \log M_1 \leq& \frac{1}{2} \log\frac{\sigma^2}{D_1} + \sqrt{\frac{1}{2n}}Q^{-1}(\epsilon_1) + O\left(\frac{\log n}{n}\right)\\
\frac{1}{n} \log M_1M_2 \leq& \frac{1}{2} \log\frac{\sigma^2}{D_2} + \sqrt{\frac{1}{2n}}Q^{-1}(\epsilon_2) + O\left(\frac{\log n}{n}\right).
\end{align}

Similarly, we can consider the case $\epsilon_1\geq \epsilon_2$. In this case, it is clear that $Q^{-1}(\epsilon_1)\leq Q^{-1}(\epsilon_2)$ and $r_1\leq r_2$. We can further divide the case into the following three cases,
\begin{enumerate}
\item  $X^n \in \cB({\bf 0}, r_1)$, i.e., $X_1^2+\cdots+X_n^2 \leq r_1^2$. In this case, both decoders can decode successfully.\\
We can find $M_1^{(1,0)}$  number of $\sqrt{nD_1}$-balls that covers $r_1$-ball where
\begin{align}
\frac{1}{n}\log M_1^{(1,0)}\leq& \frac{1}{2}\log \frac{\sigma^2}{D_1}+\sqrt{\frac{1}{2n}} Q^{-1}(\epsilon_1) +O\left(\frac{\log n}{n}\right).
\end{align} 
Similar to previous cases, we can define $\cC_1^{(1,0)}$ to be a set of the ball centers.

Also, we can cover $\sqrt{nD_1}$-ball with $M_2^{(1,0)}$ number of  $\sqrt{nD_2}$-balls where 
\begin{align}
\frac{1}{n}\log M_2^{(1,0)}\leq& \frac{1}{2}\log \frac{D_1}{D_2}+O\left(\frac{\log n}{n}\right).
\end{align}
Since $Q^{-1}(\epsilon_1)\leq Q^{-1}(\epsilon_2)$, it is clear that 
\begin{align}
\frac{1}{n}\log M_1^{(1,0)}M_2^{(1,0)}\leq \frac{1}{2}\log \frac{\sigma^2}{D_2}+\sqrt{\frac{1}{2n}} Q^{-1}(\epsilon_2) +O\left(\frac{\log n}{n}\right).
\end{align}

\item  $X^n \in \cB({\bf 0}, r_2)$ but $X^n \notin \cB({\bf 0}, r_1)$, i.e., $r_1^2<X_1^2+\cdots+X_n^2 \leq r_2^2$.\\
We will only send a message to decoder 2, and decoder 1 will declare an error. We can cover $r_2$-ball with $\tilde{M}^{(1,1)}$  number of $\sqrt{nD_2}$-balls where 
\begin{align}
\frac{1}{n}\log \tilde{M}^{(1,1)}\leq\frac{1}{2}\log \frac{\sigma^2}{D_2}+\sqrt{\frac{1}{2n}} Q^{-1}(\epsilon_2) +O\left(\frac{\log n}{n}\right).
\end{align} 
Similar to the proof of Theorem \ref{thm:AchievabilityDMS}, we can split the message $\tilde{m}^{(1,1)}\in \{1,\ldots,\tilde{M}^{(1,1)}\}$ into $(m_1^{(1,1)},m_2^{(1,1)})\in \{1,\ldots, M_1^{(1,1)}\}\times\{1,\ldots,M_2^{(1,1)}\}$ such that 
\begin{align}
& M_1^{(1,1)} M_2^{(1,1)} = \tilde{M}^{(1,1)}\\
&\frac{1}{n}\log M_1^{(1,1)}\leq\frac{1}{2}\log \frac{\sigma^2}{D_1}+\sqrt{\frac{1}{2n}} Q^{-1}(\epsilon_2) +O\left(\frac{\log n}{n}\right)\\
&\frac{1}{n}\log M_2^{(1,1)}\leq\frac{1}{2}\log \frac{D_1}{D_2}+O\left(\frac{\log n}{n}\right).
\end{align} 
Recall that the decoder 1 does not care about the reconstruction of the source, where, on the other hand, decoder 2 will get both $M_1^{(1,1)}$ and $M_2^{(1,1)}$ and will be able to reconstruct the source based on $\tilde{M}^{(1,1)}$.

\item $X^n \notin \cB({\bf 0}, r_1)$ and $X^n \notin \cB({\bf 0}, r_2)$, i.e., $r_2^2<X_1^2+\cdots+X_n^2 $.\\
We will not send any messages and both decoder will declare an error. We can think both $M_1^{(1,2)}$ and $M_2^{(1,2)}$ to be one.
\end{enumerate}
Similar to the case of $\epsilon_1<\epsilon_2$, we can combine the codebooks and get
\begin{align}
\frac{1}{n} \log M_1 \leq& \frac{1}{2} \log\frac{\sigma^2}{D_1} + \sqrt{\frac{1}{2n}}Q^{-1}(\epsilon_1) + O\left(\frac{\log n}{n}\right)\\
\frac{1}{n} \log M_1M_2 \leq& \frac{1}{2} \log\frac{\sigma^2}{D_2} + \sqrt{\frac{1}{2n}}Q^{-1}(\epsilon_2) + O\left(\frac{\log n}{n}\right).
\end{align}
This concludes the proof.

\begin{remark}\label{rem:same epsilons}
If we have $\epsilon_1=\epsilon_2=\epsilon$, radius $r_1$ and $r_2$ are the same and the proof can be simplified. In this case, an error will occur at both decoders if and only if $X_1^2+\cdots+X_n^2>r^2$ where $r=r_1=r_2$. Since both decoders share the same error events, encoding can be done successively in a simple manner and we do not have to consider the case of message splitting. More precisely, given codebook $\{(\Xh_1^n(i),\Xh_2^n(j)) : 1\leq i\leq M_1, 1\leq j \leq M_2\}$, the encoder finds $i$ such that $(1/n)\norm{X^n-\Xh_1^n(i)}_2^2\leq D_1$ and then finds $j$ such that $(1/n)\norm{X^n-\Xh_1^n(i)- \Xh_2^n(j)}_2^2 \leq D_2$. This is the key idea of Section \ref{sec:Layered Codes} where we use the successive refinement technique to construct a point-to-point source coding scheme with low complexity.
\end{remark}

%%%%%%%%%%%%%%%%%%%%%%%%%%%%%%%%%%%%%%%%%%%%%%%%
%%%%%%%%%%%%%%%%           Layered Codes                            %%%%%%%%%%%%%%%%
%%%%%%%%%%%%%%%%%%%%%%%%%%%%%%%%%%%%%%%%%%%%%%%%
\section{Layered Codes}\label{sec:Layered Codes}
We considered the successive refinement problem with two decoders so far. In this section, we show that the idea of successive refinement is also useful for point to point lossy compression where we have one encoder and one decoder. The intuition is that successive refinement coding provides a tree structure for a coding scheme which allows low encoding complexity. More precisely, if the source is successively refinable, we can add $L-1$ virtual mid-stage decoders and employ a successive refinement scheme for $L$ decoders without any (asymptotic) performance loss. For fixed $L$, this is a simple extension of successive refinement, however, we also provide a result for $L = L_n$ growing with $n$. Since the number of decoders $L$ corresponds to the level of tree and larger $L$ leads to lower complexity of the scheme, we have a great advantage in terms of complexity by taking growing $L=L_n$. 

Note that the tree structured vector quantization (TVSQ) has been extensively studied, and also has a successive approximation property. For example, in \cite{effros1994progressive}, Effros et al.\ combined pruned TVSQ with a universal noiseless coder which enables progressive transmission of sources. While this approach guarantees optimality at zero distortion,  it cannot achieve the rate-distortion function in general.

The precise problem description is the following. Let $n$ be the block length of the coding scheme. The codebook consists of $L$ sub-codebooks $(\cC_1^{(n)},\cC_2^{(n)},\cdots,\cC_L^{(n)})$ and each sub-codebook consists of $M_i$  codewords for $1\leq i\leq L$. We consider the following encoding scheme which we call \emph{layered coding}:
\begin{itemize}
\item Find $c_1\in\cC^{(n)}_1$ that minimizes some function $\psi_1(x^n,c_1)$.
\item For $i\geq 2$, given $c_1,\cdots,c_{i-1}$, find $c_i\in\cC^{(n)}_i$ that minimizes $\psi_i(x^n,c_1,c_2,\cdots,c_{i-1})$,
\end{itemize}
where $\psi_1,...,\psi_L$ are simple functions that depend on the specific implementation of the scheme. One can think of $(c_1,\ldots, c_i)$ as messages for an $i$-th (virtual) decoder. The compressed representation of the source consists of a length $L$ vector $(m_1,\cdots, m_L)$ which indicates the index of codeword from each sub-codebook. Note that the total number of codewords is $M_1\times\cdots\times M_L$ and the rate of the scheme is $R = \sum_{i=1}^L \frac{1}{n} \log M_i$. Once the decoder receives the message, it reconstructs $\hat{X}^n = \phi(m_1,\cdots,m_L)$ with some function $\phi$.
\begin{definition} 
An $(n,L,\{M_1,\cdots,M_L\},D,\epsilon)$-layered code is a coding scheme with $L$ sub-codebooks where the size of the $i$-th sub-codebook is $M_i$, and the probability of excess distortion  $\Pr{d(X^n, \hat{X}^n) >D}$ is at most $\epsilon$.
\end{definition}

Note that the definition of the layered code is exactly equal to that of the successive refinement code except the fact that the layered coding scheme only considers the distortion at the last decoder.

%%%%%%%%%%%%%%%%           Layered Coding Schemes
%%%%%%%%%%%%%%%%%%%%%%%%%%%%%%%%%%%%%%%%%%%%%%%%
%  by generalizing Theorem \ref{thm:AchievabilityDMS} and Theorem \ref{thm:AchievabilityGMS} to $L$ decoders, as stated in the theorems below
\subsection{Layered Coding Schemes}\label{subsec:Layered Coding Schemes}
We show the existence of layered coding schemes for a Gaussian source under quadratic distortion and for a binary source under Hamming distortion. For fixed $L$, it is easy to have a layered coding scheme, since sources are successively refinable in both cases and we can apply the successive refinement schemes. In this section, we generalize the result even further in two aspects. First, we consider how fast the coding rate can converge to the rate-distortion function, and provide an achievable rate including a dispersion term. Then, we allow $L$ to be a function of block length $n$, and provide a layered coding scheme for $L=L_n$ growing with $n$. Our next theorem shows an existence of a rate-distortion achieving layered coding scheme for given $n$ and $L$.
\begin{theorem}\label{thm:Layered Coding}
For i.i.d.\ Gaussian sources under quadratic distortion and i.i.d.\ binary sources under Hamming distortion, there exists a $(n,L,\{M_1,\ldots,M_L\},D,\epsilon)$-layered code such that
\begin{align}
\sum_{i=1}^L \frac{1}{n}\log M_{i}\leq R(D) + \sqrt{\frac{V(D)}{n}}Q^{-1}(\epsilon) + L\tilde{k}\frac{\log n}{n} +O\left(\frac{\log n}{n}\right)
\end{align}
for some constant $\tilde{k}$ where the $O\left({\log n}/{n}\right)$ term does not depend on $D$ or $L$.
\end{theorem}
The proof and discussion of Theorem \ref{thm:Layered Coding} are given in Section \ref{subsubsec:Gaussian source under quadratic distortion} and Section \ref{subsubsec:Binary source under Hamming distortion}. Note that $L\tilde{k} \log n /n$ is also in the class of $O(\log n/n)$ for constant $L$, however, we will also consider the case where $L=L_n$ grows with $n$. We would like to point out that the last $O(\log n/n)$ remains the same even when $L=L_n$ increases as $n$ grows.

%%%  Gaussian source under quadratic distortion
\subsubsection{ Gaussian source under quadratic distortion}\label{subsubsec:Gaussian source under quadratic distortion}
For Gaussian source under quadratic distortion, we can generalize Theorem \ref{thm:AchievabilityGMS} to the case of multiple decoders. As we mentioned in Remark \ref{rem:same epsilons}, we choose all $\epsilon_i$ to be equal to $\epsilon$.

\begin{lemma}\label{lem:GMS LC}
Let a source be i.i.d.\ Gaussian $\cN(0,\sigma^2)$ under quadratic distortion. For all $L$, there exists a\\ $(n, L, \{M_1, \cdots, M_L\}, D, \epsilon)$-layered code such that
\begin{align}
\frac{1}{n} \log M_1\leq &\frac{1}{2}\log \frac{\sigma^2}{D_1} + \sqrt{\frac{1}{2n}}Q^{-1}(\epsilon) + O\left(\frac{\log n}{n}\right)\label{eq:gaussianM1}\\
\frac{1}{n} \log M_i \leq &\frac{1}{2}\log \frac{D_{i-1}}{D_i} + 3\frac{\log n}{n} \quad\mbox{for $2\leq i\leq L$}\label{eq:gaussianMi}
\end{align}
for any $D_1>D_2>\cdots>D_L = D$ where the $O\left({\log n}/{n}\right)$ term depends on $\epsilon$ but not on $L$ or the $D_i$ values. 
\end{lemma}

The choice of $\psi_i$ and $\phi$ will be specified in the proof. The fact that the $O\left({\log n}/{n}\right)$ term is not dependent on the specific choice of $D_i$'s and $L$ is important in cases we consider later where $L$ and $D_i$ vary with $n$.

\begin{IEEEproof}
Consider the successive refinement problem with target distortions $D_1>\cdots>D_L=D$ and target excess distortion probabilities $\epsilon_1=\cdots=\epsilon_L=\epsilon$. Given sub-codebooks $\cC_1,\ldots, \cC_L$, the basic idea of the scheme is as as shown in Algorithm \ref{alg:GaussianLC}. Note that the input of the algorithm is a given sequence $x^n$ and the set of sub-codebooks $\cC_1,\ldots, \cC_L$ where the output is the collection of sub-codewords $c_{m_1},\ldots, c_{m_L}$.
\begin{algorithm}[H]
\begin{algorithmic}
\STATE Set $D_1>D_2>\cdots>D_L=D$, and let $\bx^{(0)} = x^n$.
\FOR {$i=1$ to $L$}
\STATE Find a codeword $c_{m_i}\in\cC_i$ such that $\norm{\bx^{(i-1)}-c_{m_i}}_2^2 \leq nD_i$. 
\STATE If there is no such codeword, declare an error. 
\STATE Let $\bx^{(i)} = \bx^{(i-1)}- c_{m_i}$.
\ENDFOR
\end{algorithmic}
\caption{Encoding Scheme.}
\label{alg:GaussianLC}
\end{algorithm}

We construct sub-codebooks based on Corollary \ref{cor:sphere covering}. Let $r$ be a radius such that $\Pr{X_1^2+\cdots+X_n^2 >r^2} = \epsilon$. Similar to the proof of Theorem \ref{thm:AchievabilityGMS}, we can find $M_1$ number of $\sqrt{nD_1}$-balls that covers the $r$-ball where
\begin{align}
\frac{1}{n} \log M_1\leq &\frac{1}{2}\log \frac{\sigma^2}{D_1} + \sqrt{\frac{1}{2n}}Q^{-1}(\epsilon) + O\left(\frac{\log n}{n}\right).
\end{align}
Again, the term $O\left({\log n}/{n}\right)$ only depends on $\epsilon$ where we provide the details in Appendix \ref{app:Gaussain_Oterm}. Then, for $i\geq 2$, we can cover $\sqrt{nD_{i-1}}$-ball with $M_i$ number of $\sqrt{nD_i}$-balls where
\begin{align}
\frac{1}{n} \log M_i\leq &\frac{1}{2}\log \frac{D_{i-1}}{D_i} + 3\frac{\log n}{n}.
\end{align}
 The $i$-th sub-codebook $\cC_i$ is a set of centers of $\sqrt{nD_i}$-balls, and therefore $\card{\cC_i}=M_i$.

Suppose the encoder found $c_{m_1},\cdots,c_{m_{i-1}}$ successfully, which implies $\norm{c_{m_1}+\cdots+c_{m_{i-1}} - x^n}_2^2 \leq nD_{i-1}$. In other words, $x^n$ is in the ball with radius $\sqrt{nD_{i-1}}$ where the center of the ball is at $c_{m_1}+\cdots+c_{m_{i-1}}$. Then, by construction, we can always find $c_{m_i}\in\cC_i$ such that $\psi_{i}(x^n,c_{m_1},\ldots,c_{m_{i}})=\norm{c_{m_1}+\cdots+c_{m_i} - x^n}_2^2 \leq nD_i$. We can repeat the same procedure $L$ times and find $(m_1,m_2,\ldots, m_L)$. 

The error occurs if and only if the event $X_1^2+\cdots+X_n^2>r^2$ happens at the beginning, and therefore the excess distortion probability is $\epsilon$. The reconstruction at the decoder will be $\phi(c_{m_1},\ldots,c_{m_L}) = c_{m_1}+\cdots+c_{m_L}$.
\end{IEEEproof}

The overall rate of Lemma \ref{lem:GMS LC} can be bounded by
\begin{align}
&\sum_{i=1}^L \frac{1}{n}\log M_i \\
&\leq\frac{1}{2}\log \frac{\sigma^2}{D_1}+\sqrt{\frac{1}{2n}}Q^{-1}(\epsilon)+O\left(\frac{\log n}{n}\right)+\sum_{i=2}^L \left[ \frac{1}{2}\log \frac{D_{i-1}}{D_i} + 3\frac{\log n}{n} \right]\\
&= \frac{1}{2}\log \frac{\sigma^2}{D}+\sqrt{\frac{1}{2n}}Q^{-1}(\epsilon)+3(L-1) \frac{\log n}{n} +O\left(\frac{\log n}{n}\right).\label{eq:GMS LC}
\end{align}

%%%  Binary source under Hamming distortion
\subsubsection{Binary source under Hamming distortion}\label{subsubsec:Binary source under Hamming distortion}
The next lemma provides a similar result for a binary source under Hamming distortion.
\begin{lemma}\label{lem:BMS LC}
Let the source be i.i.d.\ Bern($p$) and the distortion be measured by Hamming distortion function, where the target distortion is $D$. For large enough $n$, there is a $(n,L,\{M_1,\cdots,M_L\},D,\epsilon)$-layered code for all $L$ and $D_1>D_2>\cdots>D_L = D$ such that
\begin{align}
\frac{1}{n} \log M_1\leq & h_2(p)-h_2(D_1) + \sqrt{\frac{V(p,D)}{n}}Q^{-1}(\epsilon) + O\left(\frac{\log n}{n}\right)\label{eq:binaryM1}\\
\frac{1}{n} \log M_i \leq & h_2(D_{i-1}) -h_2(D_i) + k_3\frac{\log n}{n}\quad,\quad\mbox{for $2\leq i\leq L$}\label{eq:binaryMi}
\end{align}
where $O\left({\log n}/{n}\right)$ only depends on $\epsilon$, we denote dispersion of Bern($p$) source with $V(p,D) = p(1-p)\log^2 ((1-p)/{p})$, and a binary entropy function with $h_2(p)=-p\log p-(1-p)\log(1-p)$ and $k_3$ is a constant that does not depend on any of the variables.
\end{lemma}

\begin{IEEEproof}
Similar to the proof of Lemma \ref{lem:GMS LC}, we can consider the successive refinement problem with target distortions $D_1>\cdots>D_L=D$ and target excess distortion probabilities $\epsilon_1=\cdots=\epsilon_L=\epsilon$. The basic idea of coding is very similar to the Gaussian case. The difference is that we use Hamming instead of $l_2$ balls, and therefore we need Lemma \ref{lem:GeneralizedRefinedCovering} instead of Corollary \ref{cor:sphere covering}. A Hamming ball with radius $r$ is defined by
\begin{align}
\cB_H (r)\stackrel{\Delta}{=} \{y^n \in \{0,1\}^n : \sum_{i=1}^n y_i \leq r\}.
\end{align}

Given sub-codebooks $\cC_1,\ldots, \cC_L$, the basic idea of the achievability scheme is the following:
\begin{algorithm}[H]
\begin{algorithmic}
\STATE Set $D_1>D_2>\cdots>D_L=D$, and let $\bx^{(0)} = x^n$.
\FOR {$i=1$ to $L$}
\STATE Find the codeword $c_{m_i}\in\cC_i$ such that $d(\bx^{(i-1)},c_{m_i}) \leq D_i$. 
\STATE If there is no such codeword, declare an error. 
\STATE Let $\bx^{(i)} = \bx^{(i-1)}\oplus c_{m_i}$.
\ENDFOR
\end{algorithmic}
\caption{Enoding Scheme.}
\label{alg:BinaryLC}
\end{algorithm} 
Similar to Algorithm \ref{alg:GaussianLC}, the input of the algorithm is a given sequence $x^n$ and the set of sub-codebooks $\cC_1,\ldots, \cC_L$ where the output is the collection of sub-codewords $c_{m_1},\ldots, c_{m_L}$.

In the first stage, similar to \cite[Theorem 1]{DBLP:journals/corr/abs-1109-6310}, we can find a sub-codebook $\cC_1$ with size $M_1$ such that the excess distortion probability is smaller than $\epsilon$ and
\begin{align}
\frac{1}{n}\log M_1 \leq h(p)-h(D_1)+\sqrt{\frac{V(p,D_1)}{n}}Q^{-1}(\epsilon)+O\left(\frac{\log n}{n}\right).
\end{align}
Similar to the Gaussian case, the term $O\left({\log n}/{n}\right)$ only depends on $\epsilon$, where the detail is provided in Appendix \ref{app:Binary_Oterm}.

For $i\geq 2$ and the given type $Q$, Lemma \ref{lem:GeneralizedRefinedCovering} implies that there is $M_{Q,i}$ Hamming balls with radius $nD_i$ that covers all sequences of type $Q$ where
\begin{align}
\frac{1}{n}\log M_{Q,i}\leq& R(Q,D_i)+k_1\frac{\log n}{n}\\
=&h(Q(1)) - h(D_i) +k_1\frac{\log n}{n}.
\end{align}
Let $\cC_{Q,i}$ be a set of centers of Hamming balls with radius $nD_i$, and therefore $\card{\cC_{Q,i}}=M_{Q,i}$. The $i$-th sub-codebook $\cC_i$ is union of $\cC_{Q,i}$'s for all type $Q\in\cT(D_{i-1},D_i) \triangleq \{Q\in\cP_n(\cX):D_i<Q(1)\leq D_{i-1}\}$ and zero codeword $(0,0,0,\cdots,0)$, i.e.,
\begin{align}
\cC_i =  \{(0,\ldots,0)\}\cup\bigcup_{Q\in\cT(D_{i-1},D_i)} \cC_{Q,i}.
\end{align}
Then, we have
\begin{align}
\frac{1}{n}\log M_i &= \frac{1}{n}\log \card{\cC_i}\\
&\leq \frac{1}{n}\log \left(1+ \sum_{Q\in\cT(D_{i-1},D_i)} \card{\cC_{Q,i}}\right)\\
&\leq \frac{1}{n}\log \left(1+ (n+1) \max_{Q\in\cT(D_{i-1},D_i)} M_{Q,i}\right)\label{eq:sizeofTypes}\\
&\leq h(D_{i-1})-h(D_i)+(k_1+1)\frac{\log n}{n}
\end{align}
where \eqref{eq:sizeofTypes} is because $\card{\cT(D_{i-1},D_i)}\leq nD_{i-1}-nD_i+1$. We can set $k_3\stackrel{\Delta}{=} k_1+1$.

Suppose the encoder could find $c_{m_1},\cdots,c_{m_{i-1}}$ successfully which implies $d(c_{m_1}\oplus\cdots\oplus c_{m_{i-1}} , x^n) \leq nD_{i-1}$. In other words, $x^n$ is in the Hamming ball with radius $nD_{i-1}$ where the center of ball is at $c_{m_1}\oplus\cdots\oplus c_{m_{i-1}}$. Then, by construction, we can always find $c_{m_i}\in\cC_i$ such that $\psi_i(c_{m_1},\ldots,c_{m_i}) = d(c_{m_1}\oplus\cdots\oplus c_{m_i}, x^n)\leq nD_i$. We can repeat the same procedure $L$ times and find $(m_1,m_2,\ldots, m_L)$. 

The error occurs if and only if the first sub-codebook fails to cover the source $x^n$ at the beginning, and therefore the excess distortion probability is $\epsilon$. The reconstruction at the decoder will be $\phi(c_{m_1},\ldots,c_{m_L}) = c_{m_1}\oplus\cdots\oplus c_{m_L}$.
\end{IEEEproof}

\begin{remark}
We would like to point out that Lemma \ref{lem:BMS LC} is limited to memoryless binary sources while Theorem \ref{thm:AchievabilityDMS} holds for any discrete memoryless sources. The main difference is the operation between source symbols. More precisely, in Lemma \ref{lem:BMS LC}, the source is encoded and then the ``error" sequence (modulo 2 difference) is encoded again. Note that Hamming distortion is closely related to this operation. However, It is hard to generalize this idea to non-binary sources because there are no corresponding differences when the distortion measure is arbitrary. The modulo $|\cX|$ difference could work, but it is complex to analyse even when the distortion measure is still Hamming distortion.
\end{remark}

The overall rate of Lemma \ref{lem:BMS LC} can be bounded by
\begin{align}
\sum_{i=1}^L \frac{1}{n}\log M_i \leq& h_2(p)-h_2(D_1) +\sqrt{\frac{V(p,D_1)}{n}}Q^{-1}(\epsilon)+O\left(\frac{\log n}{n}\right)\nonumber\\
&+\sum_{i=2}^L \left[ h_2(D_{i-1}) -h_2(D_i)+ k_3\frac{\log n}{n}\right]\\
=&  h_2(p)-h_2(D)+\sqrt{\frac{V(p,D_1)}{n}}Q^{-1}(\epsilon)+k_3(L-1)\frac{\log n }{n}+O\left(\frac{\log n}{n}\right).\label{eq:BMS LC}
\end{align}

%%%%%%%%%%%%%%%%       Discussion
%%%%%%%%%%%%%%%%%%%%%%%%
\subsection{Discussion}\label{subsec:Layered_Discussion}

%%% Rate-Distortion Trade-Off
\subsubsection{Rate-Distortion Trade-Off}
In both \eqref{eq:GMS LC} and \eqref{eq:BMS LC}, it is obvious that the choice of $L$ has an important role. For simplicity, we only consider the case where $M_1=M_2=\cdots=M_L = M$, and we neglect the fact that the number of messages $M$ is an integer. We can find $M$ and $D_1>D_2>\cdots>D_L$ which satisfy \eqref{eq:gaussianM1} and \eqref{eq:gaussianMi} (or \eqref{eq:binaryM1} and \eqref{eq:binaryMi}) with equality. For example, in the Gaussian case, we can find $M$ and $D_1,\cdots,D_L$ sequentially:
\begin{align}
\frac{1}{n} \log M= &\frac{1}{2}\log \frac{\sigma^2}{D_1} + \sqrt{\frac{1}{2n}}Q^{-1}(\epsilon)+O\left(\frac{\log n}{n}\right)\\
\frac{1}{n} \log M = &\frac{1}{2}\log \frac{D_{i-1}}{D_i} + 3\frac{\log n}{n}\quad\mbox{for $2\leq i\leq L$},
\end{align}
Clearly, the number of possible reconstructions is $M^L=e^{nR}$ and the rate of the scheme is $R= ({1}/{n}) L\log M$. On the other hand, the complexity is of order $M\times L$ since the encoder is searching a right codeword over $M$ sub-codewords at each stage. Thus, for fixed rate $R$, we can say that the coding complexity (or size of codebooks) scales with $L\exp\left({nR}/{L}\right)$ which is a decreasing function of $L$. This shows that larger $L$ provides a lower complexity of the scheme. It is worth emphasizing that we can set $L=L_n$ to be increasing with $n$. This is because the bounds in both corollaries hold uniformly for all $L$.

On the other hand, in both corollaries, the overall rate can be bounded by 
\begin{align}
R(D) + \sqrt{\frac{V(D)}{n}}Q^{-1}(\epsilon) + \tilde{k}L_n\frac{\log n}{n} + O\left(\frac{\log n}{n}\right)
\end{align}
for some constant $\tilde{k}$, where we denote by $R(D)$ and $V(D)$ the rate-distortion function and the source dispersion. However, the optimum rate is given by
\begin{align}
R(D) + \sqrt{\frac{V(D)}{n}}Q^{-1}(\epsilon) + O\left(\frac{\log n}{n}\right).
\end{align}
We can see that there is a penalty term $\tilde{k}L_n(\log n/{n})$ because of using layered coding. If $L_n$ is growing too fast with $n$ in order to achieve low-complexity of the scheme, then the rate penalty term $L_n(\log n/{n})$ can be too large and we may lose (second-order) rate optimality. This shows the trade-off between the rate and complexity of the scheme. Consider the following two examples, which are valid for both the Gaussian and binary cases.
\begin{itemize}
\item If $L_n=L$ is constant, then the scheme achieves the rate-distortion and the dispersion as well, but the complexity is exponential (albeit with a smaller exponent).
\item If $L_n({\log n}/{n}) \rightarrow 0$ as $n\rightarrow \infty$, we can achieve the rate-distortion function. For example, if $L_n={n}/{\log^2 n}+1$, then the achieved rate is
\begin{align}
R = R(D) + O\left(\frac{1}{\log n}\right),
\end{align}
i.e., the scheme achieves the rate-distortion function as $n$ increases, while the coding complexity is of order $\frac{n}{\log^2 n} n^{R\log n}$. Note that the excess distortion probability $\epsilon$ is fixed. We would like to point out that the rate is near polynomial in $n$.
\item If $L_n({\log n}/{\sqrt{n}}) \rightarrow 0$ as $n\rightarrow \infty$, we can achieve the source dispersion. For example, if $L_n={\sqrt{n}}/{\log^2 n}+1$, then the achieved rate is
\begin{align}
R =& R(D)+ \sqrt{\frac{V(D)}{n}}Q^{-1}(\epsilon) + O\left(\frac{1}{\sqrt{n}\log n}\right).
\end{align}
Note that $R-R(D)$ is inversely proportional to $\sqrt{n}$ with coefficient $\sqrt{V(D)}Q^{-1}(\epsilon)$, in other words, layered coding can achieve the second order optimum rate. On the other hand, coding complexity is of order $({\sqrt{n}}/{\log^2 n}) n^{\sqrt{n}R\log n}$ which is better than the original exponential complexity.
\end{itemize}

%%% Generalized Successive Refinability
\subsubsection{Generalized Successive Refinability}
We would like to emphasize another interesting feature of layered coding. Layered coding can be viewed as a successive refinement scheme with $L$ decoders. Since our result allows $L=L_n$ to be increasing with $n$, this can be viewed as another generalized version of successive refinement. If the source is either binary or Gaussian and $\lim_{n\rightarrow \infty} L_n({\log n}/{n}) =0$, the source is successively refinable with infinitely many decoders, where the rate increment is negligible. For comparison, in the classical successive refinement result, the number of decoders is not increasing and the rate increment between neighboring decoders is strictly positive. In \cite{no2014rateless}, this property is termed \emph{infinitesimal successive refinability}, and the results here establish that Gaussian and binary sources are infinitesimally successively refinable sources (under the relevant distortion criteria). Moreover, if we further assume $\lim_{n\rightarrow \infty} L_n\frac{\log n}{\sqrt{n}} =0$, each decoder can achieve the optimum distortion including dispersion term. In this case, we can say that the binary and Gaussian sources are strongly infinitesimally successively refinable sources.

In \cite{no2014rateless}, the authors also pointed out that infinitesimal successive refinability yields another interesting property called \emph{ratelessness}. Consider a binary or Gaussian source with $\lim_{n\rightarrow \infty} L\frac{\log n}{n} =0$, where the decoder received the first few fraction of messages, i.e., $(m_1,m_2, \ldots, m_{\alpha L})$ for some $0<\alpha<1$. Based on the proof of Lemma \ref{lem:GMS LC} and Lemma \ref{lem:GMS LC}, the decoder will still be able to reconstruct the source sequence with distortion $D(\alpha R)$ which is the minimum achievable distortion at rate $\alpha R$. If we have $\lim_{n\rightarrow \infty} L_n\frac{\log n}{\sqrt{n}} =0$, an even stronger ratelessness property can be established. In this case, the decoder can achieve the optimum distortion including dispersion terms.

%%%%%%%%%%%%%%%%%%%%%%%%%%%%%%%%%%%%%%%%%%%%%%%%
%%%%%%%%%%%%%%%%           Conclusions                            %%%%%%%%%%%%%%%%
%%%%%%%%%%%%%%%%%%%%%%%%%%%%%%%%%%%%%%%%%%%%%%%%
\section{Conclusions}\label{sec:Conclusions}
We have considered the problem of successive refinement with a focus on the optimal rate including the second order dispersion term. We have proposed the concept of ``strong successive refinability" of the source and obtained a sufficient condition for it. In particular, any discrete memoryless source under Hamming distortion, or the Gaussian source under quadratic distortion are strongly successively refinable. We also show that the complexity of point-to-point source coding can be reduced using the idea of successive refinement. For binary and Gaussian sources, we characterize an achievable trade-off between rate and complexity of the scheme. We establish, for these cases, the existence of schemes which are infinitesimally successively refinable, rateless, achieve optimum dispersion, with sub-exponential complexity. Alternatively, essentially polynomial complexity is attainable if one is willing to back off from attaining the dispersion term.

\renewcommand\thesection{\Alph{section}}
\appendix
% \appendices
\renewcommand\thesection{\Alph{section}}

%%%%%%%%%%%%%%%%%%%%%%%%%%%%%%%%%%
%%%%%%% Derivative of Rate-Distortion Function %%%%%%%%%%%%
%%%%%%%%%%%%%%%%%%%%%%%%%%%%%%%%%%
\subsection{Derivative of Rate-Distortion Function}\label{app:Derivative of Rate-Distortion Function}
For fixed $D>0$, the rate-distortion function is a mapping between $\bC_m$ to $\fR$ where $\bC_m = \{(x_1,\ldots,x_m): x_i\geq 0, \forall i, \sum_{i=1}^m x_i = 1\}\subset \fR^m$. Note that the tangent space of $\bC_m$ is $(m-1)$-dimensional hyperplane that contains $\bC_m$ itself. We say $R(\cdot,D)$ is differentiable at $P=(p_1,\ldots,p_m)\in \bC_m$ if there is an extension $\tilde{R}(\cdot,D):\fR^m\ra \fR$ which is differentiable at $P$. The derivative of $R(\cdot,D)$ is defined by a derivative of its extension, i.e., 
\begin{align}
R'(P,D) \stackrel{\Delta}{=}\tilde{R}'(P,D) = \left(\frac{\partial \tilde{R}(P,D)}{\partial p_1},\frac{\partial \tilde{R}(P,D)}{\partial p_2},\ldots,\frac{\partial \tilde{R}(P,D)}{\partial p_m}\right)^T \in \fR^m
\end{align}
Since $\bC_m$ is smooth, the derivative $R'(P,D)$ is well-defined in the following sense \cite[4p]{milnor1997topology}. Let $\tilde{R}_1(\cdot,D): \fR^m\ra\fR$ be another extension of $R(\cdot,D)$, then for any $Q\in \bC_m$, we have
\begin{align}
\langle \tilde{R}_1'(P,D) , Q-P \rangle = \langle \tilde{R}'(P,D) , Q-P \rangle.\label{eq:choice of R'}
\end{align}
This implies that the derivative along its tangent plane is the same regardless of the choice of extension. This is enough to use Taylor series since
\begin{align}
R(Q,D) = R(P,D) + \langle \tilde{R}'(P,D) , Q-P \rangle + \mbox{high order terms}.
\end{align}

Now, consider the well-definedness of $V(P,D)$. For an extension $\tilde{R}(\cdot,D):\fR^m\ra\fR$, the source dispersion is defined by
\begin{align}
V(P,D) = \Var{\tilde{R}_1'(P,D)} =& \sum_{i=1}^m \left(\frac{\partial \tilde{R}_1(P,D)}{\partial p_i}\right)^2 p_i - \left(\sum_{i=1}^m \frac{\partial \tilde{R}_1(P,D)}{\partial p_i}p_i\right)^2.
\end{align}
Suppose $\tilde{R}_1(\cdot,D)$ is another extension of $R(\cdot,D)$, then \eqref{eq:choice of R'} implies that
\begin{align}
\tilde{R}_1'(P,D) = \tilde{R}'(P,D) + \alpha \1_m
\end{align}
for some $\alpha \in \fR$ where $\1_m = (1,1,\ldots,1)^T \in \fR^m$. Then, we have
\begin{align}
\Var{\tilde{R}_1'(P,D)} =& \sum_{i=1}^m \left(\frac{\partial \tilde{R}_1(P,D)}{\partial p_i}\right)^2 p_i - \left(\sum_{i=1}^m \frac{\partial \tilde{R}_1(P,D)}{\partial p_i}p_i\right)^2\\
 =& \sum_{i=1}^m \left(\frac{\partial \tilde{R}_1(P,D)}{\partial p_i}-\sum_{j=1}^m \frac{\partial \tilde{R}_1(P,D)}{\partial p_j}p_j\right)^2 p_i \\
 =& \sum_{i=1}^m \left(\frac{\partial \tilde{R}(P,D)}{\partial p_i}-\sum_{j=1}^m \frac{\partial \tilde{R}(P,D)}{\partial p_j}p_j\right)^2 p_i \\
=&\Var{\tilde{R}'(P,D)}.
\end{align}
Therefore, $\Var{R'(P,D)}$ does not depend on the particular choice of extension.

The same argument holds for $R'(P,D_1,D_2)$ and $V(P,D_1,D_2)$ as well. More precisely, for any $Q\in \bC_m$ and extensions $\tilde{R}(P,D_1,D_2)$ and $\tilde{R}_1(P,D_1,D_2)$, we have
\begin{align}
\langle \tilde{R}'(P,D_1,D_2), Q-P\rangle = \langle \tilde{R}_1'(P,D_1,D_2), Q-P\rangle.
\end{align}
Also, $\Var{R'(P,D_1,D_2)}$ does not depend on the particular choice of extension.

%%%%%%%%%%%%%%%%%%%%%%%%%%%%%%%%%%
%%%%%%% Proof of Generalized Refined Covering Lemma for Successive Refinement %%%%%%%%%%%%
%%%%%%%%%%%%%%%%%%%%%%%%%%%%%%%%%%
\subsection{Proof of Refined Covering Lemma for Successive Refinement}\label{app:Proof of Generalized Refined Covering Lemma for Successive Refinement}
\begin{figure}[h]
\centering
\begin{tikzpicture}[node distance=2.8cm, auto,>=latex', thick]
    \node [format] (src) {$X^n$};
    \node [medium, right of=src, node distance = 3cm](ch1){$W_1: \cX\rightarrow\cXh_1$};
    \node [medium, below of=ch1, node distance = 2cm](ch2){$W_2: \cX\times\cXh_1\rightarrow\cXh_2$};
    \node [format, right of=ch1, node distance = 3cm](xhat1){$\Xh_1^n$};
    \node [format, below of=xhat1, node distance = 2cm](xhat2){$\Xh_2^n$};

    \draw [->] (src) -- node {}(ch1);
    \draw [->] (src) |- node {}(ch2);
    \draw [->] (ch1) -- node {}(xhat1);
    \draw [->] (xhat1.south)  -- ++(0,-.5) -- ++(-3,0) -- node{} (ch2);
    \draw [->] (ch2) -- node{} (xhat2);
\end{tikzpicture}
\caption{Structure of Kernels}\label{fig:Structure of Kernels}
\end{figure}
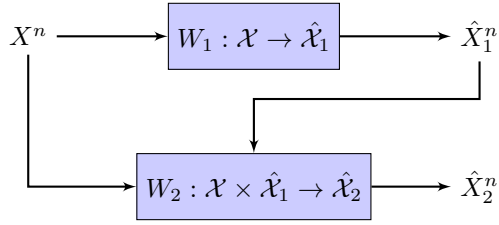

The proof is similar to the proof of \cite[Lemma 1]{kanlis1996error}, however, we have to consider vanishing terms more carefully in order to deal with source dispersions. Given type class $\cT_P$, we want to construct sets $B_1\subset \cXh_1^n$ and $B_2(\hat{x}_1^n)\subset \cXh ^n_2$ for all $\hat{x}_1^n\in B_1$ such that
\begin{align}
\cT_P\subset& \bigcup_{\hat{x}_1^n\in B_1 } \cB_1(\hat{x}_1^n,D_1),\label{eq:B condition1}\\
\cB(\hat{x}_1^n,D_1)\subset& \bigcup_{\hat{x}_2^n\in B_2(\hat{x}_1^n)} \cB_2(\hat{x}_2^n,D_2)\quad\mbox{for all $\hat x_1^n \in B_1$, } \label{eq:B condition2}
\end{align}
where $\cB_i(\hat{x}_i^n,D) = \{x^n\in\cX^n : d_i(x^n,\hat{x}_i^n)\leq D\}$ for $i\in\{1,2\}$. We construct such sets using conditional types. Let 
\begin{align}
D_1^\star =& D_1 - \frac{1}{n}\card{\cX}\cdot\card{\cXh_1}\cdot d_M\\
D_2^\star =& D_2 - \frac{1}{n}\card{\cX}\cdot\card{\cXh_1}\cdot\left(\card{\cXh_2}+1\right)d_M.
\end{align}
Then, there exist probability kernels $W_1:\cX\rightarrow \cXh_1$ and $W_2:\cX\times\cXh_1\rightarrow \cXh_2$ such that
\begin{align}
I(X;\Xh_1) =& R_1(P,D_1^\star)\\
I(X;\Xh_1,\Xh_2) =& R(P,D_1^\star,D_2^\star)
\end{align}
where the joint law of $(X,\Xh_1,\Xh_2)$ is $P\times W_1 \times W_2$ and 
\begin{align}
\E{d_1(X,\Xh_1)} =& \sum_{x,\hat{x}_1} P(x)W_1(\hat{x}_1|x)d_1(x,\hat{x}_1) \leq D_1^\star\\
\E{d_2(X,\Xh_2)} =& \sum_{x,\hat{x}_1,\hat{x}_2} P(x)W_1(\hat{x}_1|x)W_2(\hat{x}_2|x,\hat{x}_1)d_2(x,\hat{x}_2)\leq D_2^\star.
\end{align}
The structure of kernels are described in Figure \ref{fig:Structure of Kernels}.

Let $[W_1]$ and $[W_2]$ be rounded versions of $W_1$ and $W_2$ so that $n[W_1](\hat{x}_1|x)P(x)$ and $n [W_2](\hat{x}_2|x,\hat{x}_1) [W_1](\hat{x}_1|x) \\P(x)$ are integers for all $x,\hat{x}_1,\hat{x}_2$. Clearly, for all $x,\hat{x}_1,\hat{x}_2$,
\begin{align}
\card{[W_1](\hat{x}_1|x)-W_1(\hat{x}_1|x)}\leq& \frac{1}{nP(x)}\\
\card{[W_2](\hat{x}_2|x,\hat{x}_1)-W_2(\hat{x}_2|x,\hat{x}_1)}\leq& \frac{1}{n[W_1](\hat{x}_1|x)P(x)}.
\end{align}

Let $\cT_{[W_1]}(x^n)$ be the conditional type class of $[W_1]$ given $x^n$, and $\cT_{[W_2]}(x^n,\hat{x}_1^n)$ be the conditional type class of $[W_2]$ given $(x^n,\xh_1^n)$. Then, following lemma shows that $x^n,\hat{x}_1^n$ and $\hat{x}_2^n$ from those type classes satisfy distortion constraints.
\begin{lemma}\label{lem:DistortionConstraint}
For any $x^n \in \cT_P$, $\hat{x}_1^n \in\cT_{[W_1]}(x^n)$ and $\hat{x}_2^n\in\cT_{[W_2]}(x^n,\hat{x}_1^n)$, we have
\begin{align}
d_1(x^n,\hat{x}_1^n) \leq & D_1\\
d_2(x^n,\hat{x}_2^n) \leq & D_2.
\end{align}
\end{lemma}
The proof of Lemma \ref{lem:DistortionConstraint} is given in Appendix \ref{app:Proof of Lemma Distortion}.

To construct the codebook, we further let $[Q]$ be a marginalized type of $\cXh_1$ and $[V_2]$ be a marginalized kernel from $\cXh_1$ to $\cXh_2$. More precisely,
\begin{align}
[Q](\hat{x}_1) =& \sum_{x\in\cX} [W_1](\hat{x}_1|x)P(x)\\
[V_2](\hat{x}_2|\hat{x}_1) =& \frac{1}{[Q](\hat{x}_1)} \sum_{x\in\cX}  [W_2](\hat{x}_2|x,\hat{x}_1)[W_1](\hat{x}_1|x)P(x).
\end{align}
We further let $G_1 = \cT_{[Q]}$, $\tilde{G}_1(x^n) = \cT_{[W_1]}(x^n)$, $G_2(\hat{x}_1^n) = \cT_{[V_2]}(\hat{x}_1^n)$ and $\tilde{G}_2(x^n,\hat{x}_1^n) = \cT_{[W_2]}(x^n,\hat{x}_1^n)$ for all $x^n\in\cT_P$, $\hat{x}_1^n\in G_1$. It is clear that $\tilde{G}_1(x^n) \subset G_1$ and $\tilde{G}_2(x^n,\hat{x}_1^n) \subset G_2(\hat{x}_1^n)$. We generate codebook randomly based on these sets.

Let $Z^M= (Z_1,\cdots,Z_M)$ be a randomly generated codebook where $Z_1,\ldots,Z_M\in \cXh_1^n$ are i.i.d.\ random variables that has uniform distribution over $G_1$. Also, for given $Z_i=z_i$, let $\Xi_i^{N} = (\Xi_{i,1}, \cdots, \Xi_{i,N})\subset \cXh_2^n$ be i.i.d.\ random variables uniformly distributed over $G_2(z_i)$. The size of codebook $M$ and $N$ will be specified later. We denote $\cU_1(Z^M)$ the set of source words that are not covered by the codebook $Z^M$, i.e.,
\begin{align}
\cU_1(Z^M) =& \{x^n\in \cT_P : d_1(x^n,Z_i) > D_1,\mbox{ for all $1\leq i\leq M$}\}.
\end{align}
Also, for each $1 \leq i \leq M$,  let $\cU_2(\Xi_i^{N})$ be the set of source words that are covered by $Z_i$ but not covered by the codebook $\Xi_i^{N}$, i.e.,
\begin{align}
\cU_2(\Xi_i^{N}) = & \{x^n\in \cT_P : d_1(x^n,Z_i)\leq D_1, d_2(x^n,\Xi_{i,j}))> D_2,\mbox{ for all $1\leq j\leq N$}\}.
\end{align}

If we can show that $\E{\card{\cU(Z^m)\cup \left(\cup_{i=1}^M \cU_2(\Xi_i^{N})\right)}}<1 $, then we can say that there exist sets $B_1 $ and $B_2(\hat{x}_1^n)$ that satisfy \eqref{eq:B condition1} and \eqref{eq:B condition2}. This is because the random variable only gets integer values, and the fact that its expectation is less than one implies that there exists an event of the variable being equal to zero with non-zero probability, as required. We will show that the expectation can be made to be less than one, by taking $M$ and $N$ to be large enough, but not too large so that \eqref{eq:bound BPD1} and \eqref{eq:bound BPD1D2} are satisfied. Note that this argument is similar to that of \cite[Chapter 9]{csiszar2011information}. 

We begin with union bound.
\begin{align}
\E{\card{\cU_1(Z^M)\bigcup \left(\bigcup_{i=1}^m \cU_2(\Xi_i^{N})\right)}} =& \sum_{x^n \in \cT_P} \Pr{x^n \in \cU_1(Z^M)\bigcup\left(\bigcup_{i=1}^M \cU_2(\Xi_i^{N})\right)}\\
\leq & \sum_{x^n \in \cT_P} \Pr{x^n \in \cU_1(Z^M)} + \sum_{x^n \in \cT_P}  \sum_{i=1}^M\Pr{x^n\in \cU_2(\Xi_i^{N})}.\label{eq:initial upper bound}
\end{align}

We can bound the first term using type counting lemma.
\begin{align}
 \sum_{x^n \in \cT_P}\Pr{x^n\in \cU_1(Z^m)} =&  \sum_{x^n \in \cT_P}\left(1-\Pr{d_1(x^n,Z_1)\leq D_1}\right)^M\\
\leq&  \sum_{x^n \in \cT_P}\left(1-\frac{\card{\tilde{G}_1(x^n)}}{\card{G_1}}\right)^M\\
\leq&  \sum_{x^n \in \cT_P}\exp\left(-\frac{\card{\tilde{G}_1(x^n)}}{\card{G_1}}M\right)\\
\leq&  \sum_{x^n \in \cT_P}\exp\left(- (n+1)^{-\card{\cX}\cdot\card{\cXh_1}}\exp(n(H([\Xh_1]|X)-H([\Xh_1])))M\right)\label{eq:upper bound on each terms}\\
=&\card{\cT_P}\exp\left(- (n+1)^{-\card{\cX}\cdot\card{\cXh_1}}\exp(n(H([\Xh_1]|X)-H([\Xh_1])))M\right)\\
\leq& \exp(nH(P))\exp\left(- (n+1)^{-\card{\cX}\cdot\card{\cXh_1}}\exp(-nI(X;[\Xh_1]))M\right)\label{eq:upper bound on the first term}
\end{align}
where the joint law of $(X, [\Xh_1] , [\Xh_2])$ is $P \times [W_1]\times [W_2]$. Note that \eqref{eq:upper bound on each terms} is because of \eqref{eq:size of type classes} and \eqref{eq:size of conditional type classes}, while \eqref{eq:upper bound on the first term} is due to \eqref{eq:number of types}.

We can bound the second term using a similar technique.
\begin{align}
&\Pr{x^n\in \cU_2(\Xi_i^{N})}\nonumber\\
&= \Pr{d_1(x^n,Z_i)\leq D_1, d_2(x^n,\Xi_{i,j})>D_2, \forall j}\\
&=\frac{1}{\card{G_1}} \sum_{\hat{x}_1^n \in G_1} \Pr{d_1(x^n,\hat{x}_1^n)\leq D_1,d_2(x^n,\Xi_{i,j})>D_2, \forall j \suchthat Z_i = \hat{x}_1^n}\\
&=\frac{1}{\card{G_1}} \sum_{\substack{\hat{x}_1^n \in G_1\\d_1(x^n,\hat{x}_1^n)\leq D_1}} \Pr{d_2(x^n,\Xi_{i,1})>D_2 \suchthat Z_i = \hat{x}_1^n}^{N}\\
&=\frac{1}{\card{G_1}} \sum_{\substack{\hat{x}_1^n \in G_1\\d_1(x^n,\hat{x}_1^n)\leq D_1}} \exp\left(-N \frac{\card{\tilde{G}_2(x^n,\hat{x}_1^n)}}{\card{G_2(\hat{x}_1^n)}}\right)\\
&=\frac{1}{\card{G_1}} \sum_{\substack{\hat{x}_1^n \in G_1\\d_1(x^n,\hat{x}_1^n)\leq D_1}} \exp\left(-N (n+1)^{-\card{\cX}\cdot\card{\cXh_1}\cdot\card{\cXh_2}}\exp(-n(H([\Xh_2]|[\Xh_1]) - H([\Xh_2]|X,[\Xh_1])))\right).
\end{align}
Finally, we get
\begin{align}
&\sum_{x^n \in \cT_P}  \sum_{i=1}^M P\left(x^n\in \cU_2(\Xi_i^{N})\right)\nonumber\\
&\leq M\card{\cT_P}\exp\left(-N (n+1)^{-\card{\cX}\cdot\card{\cXh_1}\cdot\card{\cXh_2}}\exp(-n(H([\Xh_2]|[\Xh_1]) - H([\Xh_2]|X,[\Xh_1])))\right)\\
&\leq M|\cT_P|\exp\left(-N (n+1)^{-\card{\cX}\cdot\card{\cXh_1}\cdot\card{\cXh_2}}\exp(-nI(X;[\Xh_2]|[\Xh_1]))\right).\label{eq:upper bound on the second term}
\end{align}
We choose $M$ and $N$ that satisfy
\begin{align}
(n+1)^{\card{\cX}\cdot\card{\cXh_1}+2} \exp(nI(X;[\Xh_1]))\leq M& \leq (n+1)^{\card{\cX}\cdot\card{\cXh_1}+4} \exp(nI(X;[\Xh_1]))\\
(n+1)^{\card{\cX}\cdot\card{\cXh_1}\cdot\card{\cXh_2}+2}\exp(nI(X;[\Xh_2]|[\Xh_1])) \leq N& \leq  (n+1)^{\card{\cX}\cdot\card{\cXh_1}\cdot\card{\cXh_2}+4}\exp(nI(X;[\Xh_2]|[\Xh_1])).
\end{align}
If we apply such $M$ and $N$ to \eqref{eq:initial upper bound}, \eqref{eq:upper bound on the first term} and \eqref{eq:upper bound on the second term}, it automatically gives $\E{|\cU(Z^m)\cup \left(\cup_{i=1}^M \cU_2(\Xi_i^{N})\right)|}<1 $ for $n>\card{\cX}\cdot\card{\cXh_1}+4+H(P)+I(X;[\Xh_1])$.  Therefore, there exists sets $B_1 $ and $B_2(\hat{x}_1^n)$ that satisfies \eqref{eq:B condition1} and \eqref{eq:B condition2} where
\begin{align}
\frac{1}{n} \log |B_1 | \leq& I(X;[\Xh_1]) +  \frac{2\cdot \card{\cX}\cdot\card{\cXh_1}+8}{n}\log n\label{eq:bound B1 using [x1]}\\
\frac{1}{n}\log \left(|B_1 |\cdot|B_2(\hat{x}_1^n)|\right)\leq& I(X;[\Xh_1],[\Xh_2]) +\frac{2\cdot\card{\cX}\cdot\card{\cXh_1}\cdot\card{\cXh_2}+2\cdot\card{\cX}\cdot\card{\cXh_1}+16}{n} \log n\label{eq:bound B1B2 using [x1x2]}
\end{align}
for all $\hat{x}_1^n\in B_1 $. Note that we bound $\log(n+1)$ by $2\log n$.

Then, the following lemma bounds the gap between $I(X;\Xh_1)$ and $I(X;[\Xh_1])$ (also for $I(X;\Xh_1,\Xh_2)$ and $I(X;[\Xh_1],[\Xh_2])$) where the proof is given in Appendix \ref{app:bound the gap between mutual informations}.
\begin{lemma}\label{lem:bound the gap between mutual informations}
\begin{align}
\card{I(X;\Xh_1) - I(X;[\Xh_1])} \leq& \frac{2\card{\cX}\cdot\card{\cXh_1}}{n} \log n\\
\card{I(X;\Xh_1,\Xh_2) - I(X;[\Xh_1],[\Xh_2])} \leq& \frac{4\card{\cX}\cdot\card{\cXh_1}\cdot\card{\cXh_2}}{n} \log n.
\end{align}
\end{lemma}

With \eqref{eq:bound B1 using [x1]} and \eqref{eq:bound B1B2 using [x1x2]}, we can bound the size of $B_1$ and $B_2(\hat{x}_1^n)$'s by
\begin{align}
\frac{1}{n} \log |B_1 | \leq& I(X;\Xh_1) +  \frac{4\cdot\card{\cX}\cdot\card{\cXh_1}+8}{n}\log n \label{eq:bound B1 using x1}\\
\frac{1}{n}\log \left(|B_1 |\cdot|B_2(\hat{x}_1^n)|\right)\leq& I(X;\Xh_1,\Xh_2) +\frac{6\cdot\card{\cX}\cdot\card{\cXh_1}\cdot\card{\cXh_2}+2\cdot\card{\cX}\cdot\card{\cXh_1}+16}{n} \log   n \label{eq:bound B1B2 using x1x2}
\end{align}

Recall that we set $\Xh_1$ that satisfies $I(X;\Xh_1) = R(P,D_1^\star)$. Thus, the final step of the proof should be bounding the difference between $R(P,D_1)$ and $R(P,D_1^*)$, and also between $R(P,D_1,D_2)$ and $R(P,D_1^\star,D_2^\star)$. 
\begin{lemma}\label{lem:bound the gap between rate-distortion functions}
For large enough $n$, we have
\begin{align}
R_1(P,D_1^\star) \leq& R_1(P,D_1)+ \frac{\log n}{n}\\
R(P,D_1^*,D_2^*) \leq&  R(P,D_1,D_2)+\frac{\log n}{n}.
\end{align}
\end{lemma}
The proof is given in Appendix \ref{app:bound the gap between rate-distortion functions}

Finally, we have
\begin{align}
\frac{1}{n}\log |B_1| \leq& R_1(P,D_1)+(4\cdot\card{\cX}\cdot\card{\cXh_1}+9)\frac{\log n}{n}\\
\log\left(|B_1|\cdot|B_2(\hat{x}_1^n)| \right)\leq& R(P,D_1,D_2)+(6\cdot \card{\cX}\cdot\card{\cXh_1}\cdot\card{\cXh_2}+2\cdot\card{\cX}\cdot\card{\cXh_1}+17)\frac{\log n}{n}.
\end{align}
We can see that the coefficients of the ${\log n}/{n}$ terms are 
\begin{align}
k_1 =& 4\cdot\card{\cX}\cdot\card{\cXh_1}+9\\
k_2 =& 6\cdot\card{\cX}\cdot\card{\cXh_1}\cdot\card{\cXh_2}+2\cdot\card{\cX}\cdot\card{\cXh_1}+17
\end{align}
which are independent of the distribution $P$ and block length $n$. This concludes the proof of the lemma.

%%%%%%%%%%%%%%%%%%%%%%%%%%%%%%%%%%
%%%%%%% Proof of Corollary \ref{cor:message splitting}%%%%%%%%%%%%
%%%%%%%%%%%%%%%%%%%%%%%%%%%%%%%%%%
\subsection{Proof of Corollary \ref{cor:message splitting}}\label{app:Proof of Corollary}

By Lemma \ref{lem:GeneralizedRefinedCovering}, there exist $B_1 , \{B_2(\hat{x}_1^n)\}_{\hat{x}_1^n\in B_1}$ that successively $(D_1,D_2)$-cover $\cT_Q$ where
\begin{align}
\frac{1}{n}\log \card{B_1} \leq& R_1(Q,D_1)+k_1\frac{\log n}{n}\\
\frac{1}{n}\log\left(|B_1 |\cdot|B_2(\hat{x}_1^n)| \right)\leq& R(Q,D_1,D_2)+k_2\frac{\log n}{n}\mbox{ for all $\hat{x}_1^n \in B_1$}.
\end{align}

For simplicity, we neglect the fact that the number of messages and the size of sets are integers. Let $M_{Q,1} = e^{n\tilde{R}}$ and let $M_{Q,2}$ that satisfies $M_{Q,1}M_{Q,2} = |B_1 |\cdot \max_{\hat{x}_1^n\in B_1}|B_2(\hat{x}_1^n)|$. Then, \eqref{eq: rate bound for coding for typeq1} and \eqref{eq: rate bound for coding for typeq2} hold by definition. Then, we can find an one to one function 
\begin{align}
h: \bigcup_{\hat{x}_1^n\in B_1}  \left( \{\hat{x}_1^n\}\times  B_2(\hat{x}_1^n)\right) \rightarrow \{1,\ldots,M_{Q,1}\}\times \{1,\ldots,M_{Q,2}\}
\end{align}
such that $\hat{x}_1^n$ can be uniquely recovered based only on $m_1$ where $(m_1,m_2) = h_1( \hat{x}_1^n ,\hat{x}_2^n)$, i.e., there exists a function $\tilde{h}$ such that $\hat{x}_1^n = \tilde{h}(m_1)$. This is because $|B_1|\leq M_{Q,1}$.

For all $x^n\in \cT_Q$, there exists $\hat{x}_1^n\in B_1$ and $\hat{x}_2^n\in B_2(\hat{x}_1^n)$ such that $d_1(x^n,\hat{x}_1^n)\leq D_1$ and $d_2(x^n,\hat{x}_2^n)\leq D_2$. Let $f_{Q,1}(x^n)$ and $f_{Q,2}(x^n)$ be the first argument and the second argument of $h(\hat{x}_1^n,\hat{x}_2^n)$, respectively. Further let $g_{Q,1}(m_1) = \tilde{h}(m_1)$ and $g_{Q,2}(m_1,m_2)$ be an inverse function of $h(\cdot,\cdot)$. By construction of $B_1$ and $\{B_2(\hat{x}_1^n)\}_{\hat{x}_1^n\in B_1}$, encoder and decoder satisfies \eqref{eq:encoding for type q1} and \eqref{eq:encoding for type q2}.

Note that $M_{Q,1}$ has to be an integer, and may not be exactly equal to $e^{n\tilde{R}}$. However, we can set $({1}/{n}) \log M_{Q,1}$ to be close to $\tilde{R}$, i.e.,
\begin{align}
\tilde{R} \leq \frac{1}{n}\log  M_{Q,1} \leq \tilde{R} + \frac{\log n}{n}.
\end{align}

%%%%%%%%%%%%%%%%%%%%%%%%%%%%%%%%%%
%%%%%%% Bound $O\left(\frac{\log n}{n}\right)$ term for Gaussian case %%%%%%%%%%%%
%%%%%%%%%%%%%%%%%%%%%%%%%%%%%%%%%%
\subsection{Bound \texorpdfstring{$O\left(\frac{\log n}{n}\right)$}{} term for Gaussian case}\label{app:Gaussain_Oterm}
\begin{theorem}[Berry-Esseen Theorem \cite{berry1941accuracy}]\label{thm:Berry-Esseen}
Let $Z^n$ be i.i.d.\ random variables with $\E{Z_i}=0$, $\E{Z_i^2} = \sigma^2$ and $\E{\card{Z_i}^3} = \rho<\infty$. Let $F_n$ be the cumulative distribution function of $({\sum_{i=1}^n{Z_i}})/({\sigma\sqrt{n}})$ and $\Phi$ be the cumulative distribution function of the standard normal distribution. Then, for all $n$, 
\begin{align}
\sup_{x} \card{F_n(x) - \Phi(x)} \leq \frac{C\rho}{\sigma^3\sqrt{n}}.
\end{align}
\end{theorem}
In \cite{shevtsova2011absolute}, Shevtsova showed the optimum $C$ is smaller than $\frac{1}{2}$. 

Let $X^n$ be i.i.d.\ Gaussian random variables with zero mean and variance $\sigma^2$. Then, for $r^2>n\sigma^2$, we have
\begin{align}
\Pr{\sum_{i=1}^n X_i^2 > r^2} =& \Pr{\frac{\sum_{i=1}^n (X_i^2 -\sigma^2)}{\sqrt{2n}\sigma^2}>\frac{r^2-n\sigma^2}{\sqrt{2n}\sigma^2}}\\
\leq& Q\left(\frac{r^2-n\sigma^2}{\sqrt{2n}\sigma^2}\right) + \frac{1}{2} \frac{15\sigma^6}{2\sqrt{2n}\sigma^6}
\end{align}
where we want this probability to be smaller than $\epsilon$. Thus, we can set $r$ such that
\begin{align}
r^2 = n\sigma^2 + \sqrt{2n}\sigma^2 Q^{-1}\left(\epsilon-\frac{15}{4\sqrt{2n}}\right).
\end{align}

By Corollary \ref{cor:sphere covering}, we can cover $r$-ball with $M_1$ number of $\sqrt{nD_1}$-balls where
\begin{align}
\frac{1}{n}\log M_1 \leq& \frac{1}{2}\log\frac{r^2}{nD_1} + \frac{5}{2}\frac{\log n}{n} + \frac{1}{n}\log k_s\\
=& \frac{1}{2}\log\frac{n\sigma^2 + \sqrt{2n}\sigma^2 Q^{-1}\left(\epsilon-\frac{15}{4\sqrt{2n}}\right)}{nD_1} + \frac{5}{2}\frac{\log n}{n} + \frac{1}{n}\log k_s\\
=& \frac{1}{2}\log \frac{\sigma^2}{D_1} +\frac{1}{2}\log\left(1+\sqrt{\frac{2}{n}} Q^{-1}\left(\epsilon-\frac{15}{4\sqrt{2n}}\right)\right) + \frac{5}{2}\frac{\log n}{n} + \frac{1}{n}\log k_s\\
\leq& \frac{1}{2}\log \frac{\sigma^2}{D_1} +\frac{1}{\sqrt{2n}} Q^{-1}\left(\epsilon-\frac{15}{4\sqrt{2n}}\right) + \frac{5}{2}\frac{\log n}{n} + \frac{1}{n}\log k_s.
\end{align}
Using Taylor's expansion, one can bound $Q^{-1}\left(\epsilon-{15}/(4\sqrt{2n})\right)$ by $Q^{-1}(\epsilon) + O\left({1}/{\sqrt{n}}\right)$. Finally, we have
\begin{align}
\frac{1}{n}\log M_1 
\leq& \frac{1}{2}\log \frac{\sigma^2}{D_1} +\frac{1}{\sqrt{2n}} Q^{-1}\left(\epsilon\right) + \frac{5}{2}\frac{\log n}{n} + O\left(\frac{1}{n}\right),
\end{align}
where $O\left({1}/{n}\right)$ term does not depend on $L$ or $D_1$.

%%%%%%%%%%%%%%%%%%%%%%%%%%%%%%%%%%
%%%%%%% Bound $O\left(\frac{\log n}{n}\right)$ term for binary case %%%%%%%%%%%%
%%%%%%%%%%%%%%%%%%%%%%%%%%%%%%%%%%
\subsection{Bound \texorpdfstring{$O\left({\log n}/{n}\right)$}{} term for binary case}\label{app:Binary_Oterm}
Let $X^n$ be i.i.d.\ Bernoulli($p$) where $p<{1}/{2}$. Then, for ${1}/{2}>q>p$, we have
\begin{align}
\Pr{\sum_{i=1}^n X_i > q} =& \Pr{\frac{\sum_{i=1}^n (X_i -p)}{\sqrt{np(1-p)}}>(q-p)\sqrt{\frac{n}{p(1-p)}}}\\
\leq& Q\left((q-p)\sqrt{\frac{n}{p(1-p)}}\right) + \frac{1}{2} \frac{p}{p^{3/2}(1-p)^{3/2}\sqrt{n}}
\end{align}
where we want this probability to be smaller than $\epsilon$. Thus, we set $q$ such that
\begin{align}
q = p + \sqrt{\frac{p(1-p)}{n}} Q^{-1}\left(\epsilon - \frac{1}{2\sqrt{np(1-p)^3}}\right).
\end{align}

By Lemma \ref{lem:GeneralizedRefinedCovering}, we can cover $\cT_Q$ with $M_1$ number of $\sqrt{nD_1}$-balls where
\begin{align}
\frac{1}{n}\log M_1 \leq& h(q) - h(D_1) + k_1\frac{\log n}{n} \\
\leq& h(p) + (q-p) h'(p) - h(D_1) + k_1\frac{\log n}{n} \\
\leq& h(p) + \sqrt{\frac{p(1-p)}{n}} Q^{-1}\left(\epsilon - \frac{1}{2\sqrt{np(1-p)^3}}\right)\log \frac{1-p}{p} - h(D_1) + k_1\frac{\log n}{n} \\
=& h(p) - h(D_1) + \sqrt{\frac{V(p,D_1)}{n}} Q^{-1}\left(\epsilon - \frac{1}{2\sqrt{np(1-p)^3}}\right) + k_1\frac{\log n}{n}.
\end{align}
Using Taylor's expansion, one can bound $Q^{-1}\left(\epsilon - {1}/({2\sqrt{np(1-p)^3}})\right)$ by $Q^{-1}(\epsilon) + O\left({1}/{\sqrt{n}}\right)$. Finally, we have
\begin{align}
\frac{1}{n}\log M_1 
\leq& h(p) - h(D_1) +\frac{1}{\sqrt{n}} Q^{-1}\left(\epsilon\right) + k_1\frac{\log n}{n} + O\left(\frac{1}{n}\right),
\end{align}
where $O\left({1}/{n}\right)$ term does not depend on $L$ or $D_1$.

%%%%%%%%%%%%%%%%%%%%%%%%%%%%%%%%%%
%%%%%%% Proof of Lemma \ref{lem:DistortionConstraint} term for binary case %%%%%%%%%%%%
%%%%%%%%%%%%%%%%%%%%%%%%%%%%%%%%%%
\subsection{Proof of Lemma \ref{lem:DistortionConstraint}}\label{app:Proof of Lemma Distortion}
For any $x^n \in \cT_P$, $\hat{x}_1^n \in\cT_{[W_1]}(x^n)$ and $\hat{x}_2^n\in\cT_{[W_2]}(x^n,\hat{x}_1^n)$, we have
\begin{align}
d_1(x^n,\hat{x}_1^n) =& \sum_{x,\hat{x}_1} P(x) [W_1](\hat{x}_1|x)d_1(x,\hat{x}_1)\\
\leq& \sum_{x,\hat{x}_1} P(x) W_1(\hat{x}_1|x)d_1(x,\hat{x}_1) + \frac{1}{n} \card{\cX}\cdot\card{\cXh_1}d_M\\
\leq& D_1^\star + \frac{1}{n} \card{\cX}\cdot\card{\cXh_1}d_M\\
=& D_1.
\end{align}
Similarly, we have
\begin{align}
d_2(x^n,\hat{x}_2^n) =& \sum_{x,\hat{x}_1,\hat{x}_2} P(x) [W_1](\hat{x}_1|x)[W_2](\hat{x}_2|x,\hat{x}_1)d_2(x,\hat{x}_2)\\
\leq& \sum_{x,\hat{x}_1,\hat{x}_2} P(x) [W_1](\hat{x}_1|x)W_2(\hat{x}_2|x,\hat{x}_1)d_2(x,\hat{x}_1) + \frac{1}{n} \card{\cX}\cdot\card{\cXh_1}\cdot\card{\cXh_2}d_M\\
\leq& \sum_{x,\hat{x}_1\hat{x}_2} P(x) W_1(\hat{x}_1|x)W_2(\hat{x}_2|x,\hat{x}_1)d_2(x,\hat{x}_1) +\sum_{x,\hat{x}_1,\hat{x}_2} \frac{1}{n}W_2(\hat{x}_2|x,\hat{x}_1)d_2(x,\hat{x}_1)\nonumber\\
&+ \frac{1}{n} \card{\cX}\cdot\card{\cXh_1}\cdot\card{\cXh_2}d_M\\
\leq& D_2^\star + \frac{1}{n} \card{\cX}\cdot\card{\cXh_1}d_M+\frac{1}{n} \card{\cX}\cdot\card{\cXh_1}\cdot\card{\cXh_2}d_M\\
 =& D_2.
\end{align}

%%%%%%%%%%%%%%%%%%%%%%%%%%%%%%%%%%
%%%%%%% Proof of Lemma \ref{lem:bound the gap between mutual informations} %%%%%%%%%%%%
%%%%%%%%%%%%%%%%%%%%%%%%%%%%%%%%%%
\subsection{Proof of Lemma \ref{lem:bound the gap between mutual informations}}\label{app:bound the gap between mutual informations}

 Let $Q$ be
\begin{align}
Q(\hat{x}_1) = \sum_{x\in\cX} P(x)W_1(\hat{x}_1|x).
\end{align}
Therefore, we have
\begin{align}
\left|Q(\hat{x}_1) - [Q](\hat{x}_1)\right| =& \left|\sum_{x\in\cX} P(x) (W_1(\hat{x}_1|x) - [W_1](\hat{x}_1|x)) \right|\\
\leq & \sum_{x\in\cX} P(x) \left|W_1(\hat{x}_1|x) - [W_1](\hat{x}_1|x)\right|\\
\leq &\sum_{x\in\cX} \frac{1}{n} = \frac{\card{\cX}}{n}
\end{align}
which implies $\norm{Q-[Q]}_1 \leq {\card{\cX}\cdot\card{\cXh_1}}/{n}$. By \cite[Lemma 2.7]{csiszar2011information}, we can bound the difference between entropies:
\begin{align}
|H(\Xh_1) - H([\Xh_1])| \leq& -\frac{\card{\cX}\cdot\card{\cXh_1}}{n} \log \frac{\card{\cX}}{n}\\
\leq& \frac{\card{\cX}\cdot\card{\cXh_1}}{n} \log n.
\end{align}

Using $\tau(x) = -x\log x$, we can also bound the difference between conditional entropies:
\begin{align}
|H(\Xh_1|X) - H([\Xh_1]|X)| \leq& \sum_{x\in\cX} P(x) \left|\sum_{\hat{x}_1\in\cXh_1} \tau(W_1(\hat{x}_1|x)) - \tau([W_1](\hat{x}_1|x))\right|\\
\leq& \sum_{x\in\cX} P(x) \sum_{\hat{x}_1\in\cXh_1} \tau\left(|W_1(\hat{x}_1|x)-[W_1](\hat{x}_1|x)|\right)\\
\leq& \sum_{x\in\cX} P(x) \sum_{\hat{x}_1\in\cXh_1} \tau\left(\frac{1}{nP(x)}\right)\\
\leq& \sum_{x\in\cX} \frac{\card{\cXh_1}}{n}\log (nP(x))\label{eq:phi inequality}\\
\leq& \frac{\card{\cX}\cdot\card{\cXh_1}}{n}\log n.
\end{align}
This is because $nP(x) >3$ for all $x$. Equation \eqref{eq:phi inequality} is because $\card{\tau(x) - \tau(y)}\leq\tau(\card{x-y})$ if $\card{x-y}\leq {1}/{2}$. Finally, we get
\begin{align}
\card{I(X;\Xh_1) - I(X;[\Xh_1])} \leq& \card{H(\Xh_1)- H([\Xh_1])} + \card{H(\Xh_1|X) - H([\Xh_1]|X)}\\
\leq & \frac{\card{\cX}\cdot\card{\cXh_1}}{n} \log n +  \frac{\card{\cX}\cdot\card{\cXh_1}}{n}\log n\\
\leq &\frac{2\card{\cX}\cdot\card{\cXh_1}}{n} \log n.
\end{align}

Similarly, we can bound the difference between $I(X;\Xh_1,\Xh_2)$ and $I(X;[\Xh_1],[\Xh_2])$. Recall that $(X,\Xh_1,\Xh_2)$ has a joint law $P\times W_1\times W_2$ and $(X,[\Xh_1],[\Xh_2])$ has a joint law $P\times [W_1]\times [W_2]$.

Let $\tilde{Q}$ and $[\tilde{Q}]$ be 
\begin{align}
\tilde{Q}(\hat{x}_1,\hat{x}_2) =& \sum_x W_1(\hat{x}_1|x)W_2(\hat{x}_2|x,\hat{x}_1)P(x)\\
[\tilde{Q}](\hat{x}_1,\hat{x}_2) =& \sum_x [W_1](\hat{x}_1|x)[W_2](\hat{x}_2|x,\hat{x}_1)P(x).
\end{align}
Then, $\tilde{Q}$ and $[\tilde{Q}]$ should be similar:
\begin{align}
\card{\tilde{Q}(\hat{x}_1,\hat{x}_2) - [\tilde{Q}](\hat{x}_1,\hat{x}_2)} \leq& \sum_x P(x) \card{W_1(\hat{x}_1|x)W_2(\hat{x}_2|x,\hat{x}_1) - [W_1](\hat{x}_1|x)[W_2](\hat{x}_2|x,\hat{x}_1)}\\
\leq& \sum_x P(x) \card{W_1(\hat{x}_1|x)W_2(\hat{x}_2|x,\hat{x}_1) - [W_1](\hat{x}_1|x)W_2(\hat{x}_2|x,\hat{x}_1)}\nonumber\\
 &+ \sum_x P(x) \card{[W_1](\hat{x}_1|x)W_2(\hat{x}_2|x,\hat{x}_1) - [W_1](\hat{x}_1|x)[W_2](\hat{x}_2|x,\hat{x}_1)}\\
\leq& \sum_x \frac{1}{n}W_2(\hat{x}_2|x,\hat{x}_1) + \sum_x \frac{1}{n}\\
\leq& \frac{2\card{\cX}}{n}|
\end{align}
which implies $\norm{\tilde{Q}-[\tilde{Q}]}_1 \leq {2\card{\cX}\cdot\card{\cXh_1}\cdot\card{\cXh_2}}/{n}$. By \cite[Lemma 2.7]{csiszar2011information}, we can bound the difference between entropies
\begin{align}
\card{H(\Xh_1,\Xh_2) - H([\Xh_1],[\Xh_2])} \leq& -\frac{2\card{\cX}\cdot\card{\cXh_1}\cdot\card{\cXh_2}}{n} \log \frac{2\card{\cX}}{n}\\
\leq&\frac{2\card{\cX}\cdot\card{\cXh_1}\cdot\card{\cXh_2}}{n}\log n.\label{eq:Bound entropy x1x2}
\end{align}

Note that
\begin{align}
& \card{W_1(\hat{x}_1|x)W_2(\hat{x}_2|x,\hat{x}_1) - [W_1](\hat{x}_1|x)[W_2](\hat{x}_2|x,\hat{x}_1)} \nonumber\\
&\leq \card{W_1(\hat{x}_1|x)W_2(\hat{x}_2|x,\hat{x}_1) - [W_1](\hat{x}_1|x)W_2(\hat{x}_2|x,\hat{x}_1)}\nonumber\\
&+\card{[W_1](\hat{x}_1|x)W_2(\hat{x}_2|x,\hat{x}_1) - [W_1](\hat{x}_1|x)[W_2](\hat{x}_2|x,\hat{x}_1)}\\
&\leq  \frac{1}{nP(x)}W_2(\hat{x}_2|x,\hat{x}_1) +\frac{1}{nP(x)}\\
&\leq \frac{2}{nP(x)}
\end{align}
Since we assumed that $nP(x) >3$, we have
\begin{align}
&\card{H(\Xh_1,\Xh_2|X) - H([\Xh_1],[\Xh_2]|X)}\nonumber\\
&\leq \sum_x P(x) \card{\sum_{\hat{x}_1,\hat{x}_2} \tau(W_1(\hat{x}_1|x)W_2(\hat{x}_2|x,\hat{x}_1)) - \tau([W_1](\hat{x}_1|x)[W_2](\hat{x}_2|x,\hat{x}_1))}\\
&\leq \sum_x P(x) \sum_{\hat{x}_1,\hat{x}_2} \tau(\card{ W_1(\hat{x}_1|x)W_2(\hat{x}_2|x,\hat{x}_1)-[W_1](\hat{x}_1|x)[W_2](\hat{x}_2|x,\hat{x}_1)})\\
&\leq \sum_x P(x) \sum_{\hat{x}_1,\hat{x}_2} \tau\left(\frac{2}{nP(x)}\right)\\
&\leq \sum_x \frac{2\card{\cXh_1}\cdot\card{\cXh_2}}{n}\log \frac{nP(x)}{2}\\
&\leq \frac{2\card{\cX}\cdot\card{\cXh_1}\cdot\card{\cXh_2}}{n}\log n.\label{eq:Bound cond entropy x1x2 given x}
\end{align}

Using \eqref{eq:Bound entropy x1x2} and \eqref{eq:Bound cond entropy x1x2 given x}, we can bound the gap between mutual informations:
\begin{align}
\card{I(X;\Xh_1,\Xh_2) - I(X;[\Xh_1],[\Xh_2])} \leq& \card{H(\Xh_1,\Xh_2)- H([\Xh_1],[\Xh_2])} + \card{H(\Xh_1,\Xh_2|X) - H([\Xh_1],[\Xh_2]|X)}\\
\leq&\frac{2\card{\cX}\cdot\card{\cXh_1}\cdot\card{\cXh_2}}{n}\log n + \frac{2\card{\cX}\cdot\card{\cXh_1}\cdot\card{\cXh_2}}{n}\log n\\
\leq &\frac{4\card{\cX}\cdot\card{\cXh_1}\cdot\card{\cXh_2}}{n} \log n.
\end{align}

%%%%%%%%%%%%%%%%%%%%%%%%%%%%%%%%%%
%%%%%%% Proof of Lemma \ref{lem:bound the gap between rate-distortion functions} %%%%%%%%%%%%
%%%%%%%%%%%%%%%%%%%%%%%%%%%%%%%%%%
\subsection{Proof of Lemma \ref{lem:bound the gap between rate-distortion functions}}\label{app:bound the gap between rate-distortion functions}

We know that $D^\star_1 = D_1-\card{\cX}\cdot|\cXh_1|d_M/n$. Using the convexity and monotonicity properties of the rate-distortion function, we find an upper bound on the difference between $R(P,D_1^\star)$ and $R_1(P,D_1)$:
\begin{align}
-\frac{R_1(P,D^\star_1) - R_1(P,D_1)}{D_1^\star-D_1}\leq& \frac{R_1(P,0)}{D_1}\\
\leq& \frac{\log \card{\cX}}{D}.
\end{align}
Therefore, we can bound $R_1(P,D_1^\star)$ using $R_1(P,D_1)$:
\begin{align}
R_1(P,D^\star_1) \leq& R_1(P,D_1)+\card{\cX}\cdot\card{\cXh_1}d_M\frac{\log \card{\cX}}{nD_1}\\
\leq& R_1(P,D_1)+ \frac{\log n}{n}
\end{align}
for large enough $n$. Similarly, by the mean value theorem, there exists a $c$ such that for large enough $n$,
\begin{align}
R(P,D_1^*,D_2^*) -R(P,D_1,D_2) \leq& \left<\nabla R(P,D'_1,D'_2),  (D_1-D_1^*,D_2-D_2^*)\right>\\
\leq& \frac{\log n}{n}
\end{align}
where $D'_1 = cD_1+(1-c)D_1^*$, $D'_2 = cD_2+(1-c)D_2^*$.

%%%%%%%%%%%%%%          Bibliography     

\bibliographystyle{IEEEtran}
%\bibliography{../../AlbertRef}
% for arxiv submission

\bibliography{../../AlbertRef}

\begin{thebibliography}{10}
\providecommand{\url}[1]{#1}
\csname url@samestyle\endcsname
\providecommand{\newblock}{\relax}
\providecommand{\bibinfo}[2]{#2}
\providecommand{\BIBentrySTDinterwordspacing}{\spaceskip=0pt\relax}
\providecommand{\BIBentryALTinterwordstretchfactor}{4}
\providecommand{\BIBentryALTinterwordspacing}{\spaceskip=\fontdimen2\font plus
\BIBentryALTinterwordstretchfactor\fontdimen3\font minus
  \fontdimen4\font\relax}
\providecommand{\BIBforeignlanguage}[2]{{%
\expandafter\ifx\csname l@#1\endcsname\relax
\typeout{** WARNING: IEEEtran.bst: No hyphenation pattern has been}%
\typeout{** loaded for the language `#1'. Using the pattern for}%
\typeout{** the default language instead.}%
\else
\language=\csname l@#1\endcsname
\fi
#2}}
\providecommand{\BIBdecl}{\relax}
\BIBdecl

\bibitem{koshelev1980hierarchical}
V.~Koshelev, ``Hierarchical coding of discrete sources,'' \emph{Problemy
  peredachi informatsii}, vol.~16, no.~3, pp. 31--49, 1980.

\bibitem{koshelev1981estimation}
------, ``Estimation of mean error for a discrete successive-approximation
  scheme,'' \emph{Problemy Peredachi Informatsii}, vol.~17, no.~3, pp. 20--33,
  1981.

\bibitem{equitz1991successive}
W.~H. Equitz and T.~M. Cover, ``Successive refinement of information,''
  \emph{Information Theory, IEEE Transactions on}, vol.~37, no.~2, pp.
  269--275, 1991.

\bibitem{rimoldi1994successive}
B.~Rimoldi, ``Successive refinement of information: Characterization of the
  achievable rates,'' \emph{Information Theory, IEEE Transactions on}, vol.~40,
  no.~1, pp. 253--259, 1994.

\bibitem{kanlis1996error}
A.~Kanlis and P.~Narayan, ``Error exponents for successive refinement by
  partitioning,'' \emph{Information Theory, IEEE Transactions on}, vol.~42,
  no.~1, pp. 275--282, 1996.

\bibitem{tuncel2003error}
E.~Tuncel and K.~Rose, ``Error exponents in scalable source coding,''
  \emph{Information Theory, IEEE Transactions on}, vol.~49, no.~1, pp.
  289--296, 2003.

\bibitem{marton1974error}
K.~Marton, ``Error exponent for source coding with a fidelity criterion,''
  \emph{Information Theory, IEEE Transactions on}, vol.~20, no.~2, pp.
  197--199, 1974.

\bibitem{ingber2011dispersion}
A.~Ingber and Y.~Kochman, ``The dispersion of lossy source coding,'' in
  \emph{Data Compression Conference (DCC)}.\hskip 1em plus 0.5em minus
  0.4em\relax IEEE, 2011, pp. 53--62.

\bibitem{kostina2012fixed}
V.~Kostina and S.~Verd{\'u}, ``Fixed-length lossy compression in the finite
  blocklength regime,'' \emph{Information Theory, IEEE Transactions on},
  vol.~58, no.~6, pp. 3309--3338, 2012.

\bibitem{gupta2008rate}
A.~Gupta, S.~Verdu, and T.~Weissman, ``Rate-distortion in near-linear time,''
  in \emph{Information Theory, 2008. ISIT 2008. IEEE International Symposium
  on}.\hskip 1em plus 0.5em minus 0.4em\relax IEEE, 2008, pp. 847--851.

\bibitem{korada2010polar}
S.~B. Korada and R.~L. Urbanke, ``Polar codes are optimal for lossy source
  coding,'' \emph{Information Theory, IEEE Transactions on}, vol.~56, no.~4,
  pp. 1751--1768, 2010.

\bibitem{venkataramanan2014lossy}
R.~Venkataramanan, T.~Sarkar, and S.~Tatikonda, ``Lossy compression via sparse
  linear regression: Computationally efficient encoding and decoding,''
  \emph{Information Theory, IEEE Transactions on}, vol.~60, no.~6, pp.
  3265--3278, June 2014.

\bibitem{no2014rateless}
A.~No and T.~Weissman, ``Rateless lossy compression via the extremes,''
  \emph{arXiv preprint arXiv:1406.6730}, 2014.

\bibitem{yang1999redundancy}
E.-h. Yang and Z.~Zhang, ``On the redundancy of lossy source coding with
  abstract alphabets,'' \emph{Information Theory, IEEE Transactions on},
  vol.~45, no.~4, pp. 1092--1110, 1999.

\bibitem{zhang1997redundancy}
Z.~Zhang, E.-H. Yang, and V.~K. Wei, ``The redundancy of source coding with a
  fidelity criterion. 1. known statistics,'' \emph{Information Theory, IEEE
  Transactions on}, vol.~43, no.~1, pp. 71--91, 1997.

\bibitem{DBLP:journals/corr/abs-1109-6310}
\BIBentryALTinterwordspacing
D.~Wang, A.~Ingber, and Y.~Kochman, ``The dispersion of joint source-channel
  coding,'' \emph{CoRR}, vol. abs/1109.6310, 2011. [Online]. Available:
  \url{http://arxiv.org/abs/1109.6310}
\BIBentrySTDinterwordspacing

\bibitem{chow1997failure}
J.~Chow and T.~Berger, ``Failure of successive refinement for symmetric
  {G}aussian mixtures,'' \emph{Information Theory, IEEE Transactions on},
  vol.~43, no.~1, pp. 350--352, 1997.

\bibitem{erokhin1958varepsilon}
V.~Erokhin, ``$\varepsilon$-entropy of a discrete random variable,''
  \emph{Theory of Probability \& Its Applications}, vol.~3, no.~1, pp. 97--100,
  1958.

\bibitem{csiszar2011information}
I.~Csiszar and J.~K{\"o}rner, \emph{Information theory: coding theorems for
  discrete memoryless systems}.\hskip 1em plus 0.5em minus 0.4em\relax
  Cambridge University Press, 2011.

\bibitem{rogers1963covering}
C.~Rogers, ``Covering a sphere with spheres,'' \emph{Mathematika}, vol.~10,
  no.~02, pp. 157--164, 1963.

\bibitem{effros1994progressive}
M.~Effros, P.~A. Chou, E.~A. Riskin, and R.~M. Gray, ``A progressive universal
  noiseless coder,'' \emph{Information Theory, IEEE Transactions on}, vol.~40,
  no.~1, pp. 108--117, 1994.

\bibitem{milnor1997topology}
J.~W. Milnor, \emph{Topology from the differentiable viewpoint}.\hskip 1em plus
  0.5em minus 0.4em\relax Princeton University Press, 1997.

\bibitem{berry1941accuracy}
A.~C. Berry, ``The accuracy of the {G}aussian approximation to the sum of
  independent variates,'' \emph{Transactions of the american mathematical
  society}, vol.~49, no.~1, pp. 122--136, 1941.

\bibitem{shevtsova2011absolute}
I.~Shevtsova, ``On the absolute constants in the {B}erry-{E}sseen type
  inequalities for identically distributed summands,'' \emph{arXiv preprint
  arXiv:1111.6554}, 2011.

\end{thebibliography}

\end{document}